\documentclass[preprint]{aastex}

\def\la{\mathrel{\hbox{\rlap{\hbox{\lower4pt\hbox{$\sim$}}}\hbox{$<$}}}}
\def\ga{\mathrel{\hbox{\rlap{\hbox{\lower4pt\hbox{$\sim$}}}\hbox{$>$}}}}

\newcommand{\be}{\begin{eqnarray}}
\newcommand{\ee}{\end{eqnarray}}

\newcommand{\msol}{\ifmmode{{\rm M}_\odot}\else{M$_\odot$}\fi}
\newcommand{\foe}{\ifmmode{10^{51}}\else{$10^{51}$}\fi}

\newcommand{\xni}{\ifmmode{{\rm X}_{\rm Ni}}\else{X$_{\rm Ni}$}\fi}

\def\Teff{\ifmmode{T_{\rm eff}}\else{\hbox{$T_{\rm eff}$} }\fi}
\def\Rzero{\ifmmode{R_0}\else{\hbox{$R_0$} }\fi}

\def\SP2{{\tt IBM SP2}}

\def\PC2{{\tt PC$^2$}}

\def\inu{\ifmmode{I_{\nu}}\else{\hbox{$I_{\nu}$} }\fi}
\def\snu{\ifmmode{S_{\nu}}\else{\hbox{$S_{\nu}$} }\fi}
\def\jnu{\ifmmode{J_{\nu}}\else{\hbox{$J_{\nu}$} }\fi}

\def\etal{et al.}
\def\fep{\ifmmode{{\rm Fe II}}\else\hbox{Fe~II }\fi}

  at 12.0 true pt 

\def\etal{{et al}}

\def\phx{{\tt PHOENIX}}

\def\etal{{et al}}

\def\phx{{\tt PHOENIX}}

\def\hitran{{\tt HITRAN92}}

\def\b{\beta}

\def\rout{\ifmmode{r_{\rm out}}\else\hbox{$r_{\rm out}$}\fi}
\def\tmax{\ifmmode{\tau_{\rm max}}\else\hbox{$\tau_{\rm max}$}\fi}
\def\tstd{\ifmmode{\tau_{\rm std}}\else\hbox{$\tau_{\rm std}$}\fi}
\def\vmax{\ifmmode{v_{\rm max}}\else\hbox{$v_{\rm max}$}\fi}
\def\muE{\ifmmode{\mu_{\rm E}}\else\hbox{$\mu_{\rm E}$}\fi} 
\def\pE{\ifmmode{p_{\rm E}}\else\hbox{$p_{\rm E}$}\fi} 
\def\bmax{\ifmmode{\b_{\rm max}}\else\hbox{$\b_{\rm max}$}\fi}
\def\kms{\hbox{$\,$km$\,$s$^{-1}$}}

\def\Teff{\hbox{$\,T_{\rm eff}$} }

\def\alog#1{\times 10^{#1}}

\def\rout{\hbox{$r_{\rm out}$} }

\def\chistd{\ifmmode{\chi_{\rm std}}\else\hbox{$\chi_{\rm std}$}\fi}

\def\msol{$M_\odot$}
\def\foe{10^{51}}

\def\xni{{\rm X}_{\rm Ni}}

\def\teff{{\rm T}_{\rm eff}}

\def\lstar{\ifmmode{\Lambda^*}\else\hbox{$\Lambda^*$}\fi} 
\def\Rop{\ifmmode{[R_{ij}]}\else\hbox{$[R_{ij}]$}\fi}

\def\Rji{\ifmmode{[R_{ji}]}\else\hbox{$[R_{ji}]$}\fi}
\def\Rstar{\ifmmode{[R_{ij}^*]}\else\hbox{$[R_{ij}^*]$}\fi}

\def\Rjistar{\ifmmode{[R_{ji}^*]}\else\hbox{$[R_{ji}^*]$}\fi}
\def\DRji{\ifmmode{[\Delta R_{ji}]}\else\hbox{$[\Delta R_{ji}]$}\fi}
\def\DRij{\ifmmode{[\Delta R_{ij}]}\else\hbox{$[\Delta R_{ij}]$}\fi}

\def\ns{\ifmmode{N_{\rm s}}          
        \else\hbox{$N_{\rm s}$}\fi}

\def\mat#1{{\bf #1}}     
\def\vek#1{{#1}}         

\newcount\eqcount
\def
  \nummer{
    \global\advance\eqcount by 1
    (\the\eqcount)
  }

\def
  \numadv{
    \global\advance\eqcount by 1
  }

\def
   \numout#1{
     (\the\eqcount #1)
  }

\def\ivek#1#2{\ifmmode{\vek{I}^{#1}_{#2}}
        \else\hbox{$\vek{I}^{#1}_{#2}$}\fi}

\def\tmat#1#2{\ifmmode{\mat{t}^{#1}_{#2}}
        \else\hbox{$\mat{t}^{#1}_{#2}$}\fi}
\def\rmat#1#2{\ifmmode{\mat{r}^{#1}_{#2}}
        \else\hbox{$\mat{r}^{#1}_{#2}$}\fi}
\def\bvek#1#2{\ifmmode{\beta^{#1}_{#2}}
        \else\hbox{$\beta^{#1}_{#2}$}\fi}

\def\lp{\ifmmode{\lambda^+_\tau}           
        \else\hbox{$\lambda^+_\tau$}\fi}
\def\lm{\ifmmode\lambda^-_\tau             
        \else\hbox{$\lambda^-_\tau$}\fi}

\def\teff{{\rm T}_{\rm eff}}

\def\hitran{{\tt HITRAN92}}

\usepackage{psfig,natbib}

\shortauthors{Allard et al.}
\shorttitle{Brown Dwarfs Models}

\begin{document}
\bibliographystyle{apj}

\title{The Limiting Effects of Dust in Brown Dwarf Model Atmospheres}

\author{France Allard}
\affil{Centre de Recherche Astronomique de Lyon (CRAL),
Ecole Normale Sup\'erieure de Lyon, Lyon, Cedex 07, 69364 France\\
E-Mail: {\tt fallard@ens-lyon.fr}}

\author{Peter H. Hauschildt}
\affil{Dept.\ of Physics and Astronomy \& Centre for Simulational Physics, 
University of Georgia, Athens, GA 30602-2451\\
Email: {\tt yeti@hal.physast.uga.edu}}

\author{David R. Alexander, Akemi Tamanai}
\affil{Dept.\ of Physics, Wichita State University,
Wichita, KS 67260-0032\\
E-Mail: {\tt dra@twsuvm.uc.twsu.edu}}

\author{and Andreas Schweitzer}
\affil{Dept.\ of Physics and Astronomy,
University of Georgia, Athens, GA 30602-2451\\
Email: {\tt andy@physast.uga.edu}}

\begin{abstract}
We present  opacity sampling model atmospheres,  synthetic spectra and
colors for brown  dwarfs and very low mass stars  in two limiting case
of dust  grain formation:  1) inefficient gravitational  settling i.e.
the  dust  is  distributed   according  to  the  chemical  equilibrium
predictions, 2) efficient gravitational  settling i.e.  the dust forms
and depletes refractory elements from  the gas, but their opacity does
not affect the thermal structure.  The models include the formation of
over 600  gas phase  species, and 1000  liquids and crystals,  and the
opacities  of   30  different  types  of   grains  including  corundum
(Al$_2$O$_3$),  the  magnesium  aluminum spinel  MgAl$_2$O$_4$,  iron,
enstatite  (MgSiO$_3$), forsterite (Mg$_2$SiO$_4$),  amorphous carbon,
SiC, and  a number of calcium  silicates.  The models  extend from the
beginning  of the grain  formation regime  well into  the condensation
regime of water ice ($\teff=  3000 - 100$~K) and encompasses the range
of $\log g= 2.5 - 6.0$ at solar metallicity.

We find  that silicate  dust grains can  form abundantly in  the outer
atmospheric layers  of red and  brown dwarfs with spectral  type later
than M8.  The  greenhouse effects of dust opacities  provide a natural
explanation for  the peculiarly red spectroscopic  distribution of the
latest M dwarfs  and young brown dwarfs.  The  grainless (Cond) models
on the other hand, correspond  closely to methane brown dwarfs such as
Gliese 229B.   We also recover that  the $\lambda$5891,5897\AA\ Na~I~D
and  $\lambda$7687,7701\AA\ K~I  resonance doublets  plays  a critical
role in T dwarfs where their red wing define the pseudo-continuum from
the $I$ to the $Z$ bandpass.

\end{abstract}


\section{Introduction}\label{intro}

The discovery of the first unambiguous evolved brown dwarf Gliese 229B
\cite[]{1995Sci...270.1478O,1995Natur.378..463N},  the confirmation of
the  existence of  young brown  dwarfs  in the  Pleiades open  cluster
\cite[]{1996ApJ...469L..53R,Basri96,1997ApJ...491L..81Z},    and   the
detection        of         dozens        of        others        from
photometric~\cite[]{1997A&A...327L..25D,kirk99,Strauss99,Tsvetanov00}
and proper motion surveys \cite[]{kelu1} has restricted the intriguing
gap  between stars  and planets.   In  fact, brown  dwarfs are  bright
enough  to be  easily detected  in standard  bandpasses  ($ZJHK$) from
ground-based  facilities.   This  is  now understood  as  the  natural
consequence  of  strong absorption  bands  of  H$_2$O  and H$_2$  that
depress the  infrared flux in  favor of the  near-infrared bandpasses,
far  from  any black  body  distributions.   This  effect was  already
apparent  from the  results of  our  model atmospheres~\cite[]{MDpap},
which  were  later extended  deeper  into  the  temperature regime  of
evolved brown dwarfs by \cite{gl229b}.  But as brown dwarf discoveries
unfold, several questions  arise.  Why do brown dwarfs  appear to form
two distinct subgroups:  1) hotter red objects just  below the stellar
main sequence,  and 2)  much cooler and  blue methane dwarfs?   Is the
apparent  gap  between  these  groups real?   Are  different  physical
processes involved in their atmospheres beyond the change of effective
temperature?

From  previous   model  atmospheres,  it   immediately  appeared  more
difficult to  explain the  behavior of the  hotter brown  dwarfs, than
that  of the methane  dwarfs in  which water  vapor and  methane bands
naturally matched the  predictions of models (\cite[]{gl229b,tsuji96b}
and  \cite[]{marley96})  models   within  the  accuracy  of  available
molecular   absorption  profiles.   Even   the  most   detailed  model
atmospheres had  failed to reproduce accurately  the spectroscopic and
photometric properties of  red dwarfs later then about  M6: all models
were systematically  too blue  by as much  as a magnitude  in standard
infrared colors ($V-K$, $I-K$,  $J-K$).  \cite{araa} has reviewed this
situation  and the  atmosphere  modeling  at the  bottom  of the  main
sequence.  The reasons  for these discrepancies was the  onset of dust
grain  formation.  As  mass  decreases along  the  main sequence,  the
latest-type  red  dwarfs bear  outer  atmospheric  layers which  reach
temperatures well below 1800K,  favoring the formation of dust grains.
While it  has been long suspected  that grains could  form under these
conditions  \cite[]{lunine89}, the  inclusion of  such  computation to
non-grey  model atmosphere calculations  had to  wait until  --- using
Gibbs  free energies of  formation \cite[]{1993ApJ...406..158B}  and a
simple  sphere approximation  for the  Mie opacity  of the  grains ---
\cite{tsuji96b} published their exploratory non-grey dusty brown dwarf
models.   They explored  the formation  of three  dust  grain species:
Al$_2$O$_3$, Fe, and MgSiO$_3$, and found corundum (Al$_2$O$_3$) to be
a very  abundant and powerful  continuous absorber in red  dwarfs with
spectral  type  later  than  M8,  while cooler  methane  brown  dwarfs
appeared comparatively grainless.  But their models were based on band
models  for molecular opacities  and could  not reproduce  the optical
spectral  distributions and  several photometric  properties  of brown
dwarfs.

Recently,  \cite{tsuji99} and \cite{Ack00}  have explored  models with
finite radial  extention of silicate clouds to  address the systematic
difference between early L and late  T type brown dwarfs.  While it is
pertinent   to  explore  such   processes,  several   parameters  must
inevitably  be used  to  characterize them,  and  moreover, these  are
likely time-dependant processes \cite[e.g. I-band variability has been
recently  detected  in young  brown  dwarfs]{BJ01}.   So  while we  do
believe  that dust  diffusion (referred  to herafter  as gravitational
settling) contributes  in reducing the  radial extention of  clouds in
these atmospheres, we feel that, as in all unsolved physical problems,
it  is important  to explore  carefully the  limiting cases.   In this
paper, we  present therefore our  model calculations for  two opposite
limiting  cases  of  dust  content  in brown  dwarfs  atmospheres:  1)
inefficient  gravitational  settling  i.e.   the dust  is  distributed
according  to  chemical equilibrium  predictions,  which provides  the
maximum  impact  of dust  upon  the  brown  dwarf properties,  and  2)
efficient  gravitational settling  i.e.  the  dust forms  and depletes
refractory elements  from the gas,  but their opacity does  not affect
the  thermal  structure, hence  a  minimal  effect  upon brown  dwarfs
properties.

These models bring substantial  improvements upon the previous similar
work of \cite{tsuji96a}  in the number of grains  included both in the
chemical  equilibrium and  in the  opacity database,  as well  as with
regards to the molecular  opacities.  In Section \ref{eos} we describe
how hundreds  of grain species  are now included  self-consistently to
the chemical equilibrium  calculation to allow us to  identify the hot
condensates that are  most abundant in the atmospheres  of late-type M
dwarfs  and brown  dwarfs.  In  Section \ref{clouds}  we  describe our
treatment of the grain opacities.  And Section \ref{molec} we describe
the detailed  opacity sampling model atmospheres to  which these grain
formation  and  opacities are  incorporated.   Section \ref{struc}  is
reserved for  the discussion of the atmospheric  structures, while the
effects of  grains on the spectroscopic and  photometric properties of
brown dwarfs are discussed in Section \ref{spec} and \ref{color}.

\section{\label{eos}Condensation and the Chemical Equilibrium}

The chemical equilibrium (hereafter CE) of the atmosphere code Phoenix
already  solves simultaneously  for 40  elements with  usually 2  to 6
ionization stages per element,  and 600 molecular species relevant for
oxygen-rich  ideal  gas  compositions.   This CE  has  been  gradually
updated since \cite{MDpap} with additional molecular species using the
polynomial  partition   functions  by  \cite{irwin88}   and  \cite[][,
hereafter  SH90]{Sharp&Huebner90}.   The  molecular ions  TiO$^+$  and
ZrO$^+$, which have  been found to be important in  the balance of TiO
and  ZrO  when departures  from  local  thermodynamic equilibrium  are
present \cite[]{tipap}, have been  added using the partition functions
by  \cite{TiOplus}.   The  ions   H$_3^+$  (and  H$_2^+$),  which  are
important electron donors in  low metallicity subdwarf stars and white
dwarfs \cite[see]{1994ApJ...424..333S},  has been added to  our CE and
opacity database by  using the partition function by  \cite{h3p} and a
list of 3 million transitions by \cite{h3pl}.

To include dust  grains, we have expanded the system  of CE to include
the complete series of over  1000 liquids and crystals also studied by
SH90.  For each grain or liquid species, we followed the prescriptions
of \cite{Grossman72}  and used Gibbs free energy  of formation $\Delta
G(T)$, drawn from the JANAF 1986 database~\cite{janaf}, to compute the
so-called equilibrium  pressures $P_{eq} =  \exp^{-\Delta G(T)/RT}$ of
the  grains, where  $R$ is  the  gas constant  and $T$  the local  gas
temperature.   $P_{eq}$  was then  compared  to  the pressure,  $P_c$,
obtained from  the Guldberg law  of mass-action (i.e.  the  product of
the  pressures of  the  constituting elements).   The  abundance of  a
condensed  species was  then  determined by  the  condition that  this
species be  in equilibrium with the  gas phase, $P_c  \ge P_{eq}$. For
corundum,

$$ P(Al_2O_3) = {P(Al)^2 P(O)^3 \over P_{eq}} P(Al_2O_3)^{max};$$

$$\quad for \,\, P_{eq} \le P(Al)^2 P(O)^3$$

\noindent
where $max$  refers to the  maximum concentration of grain  cores (one
core of corundum = one Al$_2$O$_3$ unit) given the conservation of the
cores  of  each  elements.   The   complete  CE  was  then  solved  by
Newton-Raphson  iteration  of  the  equation system  \cite[see  also]{
allard90,MDpap} until  the error relative  to the gas pressure  of all
partial  pressures was  below $-6$  dex.  The  computations  were then
performed and  tabulated in  a $[P,T]-$plane encompassing  largely the
conditions prevailing in low mass stars  and brown dwarfs ($P = -4$ to
12 and  $T = 15000$ to  10K), using a solar  mix \cite[]{abund} except
for lithium  (meteoretic abundance, same source).  The  CE tables were
then interpolated in the construction of the model atmospheres.

The  CE therefore  accounts, self-consistently,  for the  depletion of
refractory elements as a function  of the gas temperature and pressure
conditions in the model atmospheres.  However, it should be clear that
thermodynamic equilibrium studies only tell us what can be formed, not
what is  formed.  And  the results  are only as  certain as  the JANAF
equilibrium constants on  which they are based.  In  favor of CE, note
however  that  \cite{LF98},  who  used thermodynamics,  were  able  to
explain the abundance patterns  of various trace elements dissolved in
carbide   stardust.   We   therefore  feel   confident  that   our  CE
calculations allow  us at least  to reproduce, in  average, adequately
the  limiting thermal  and spectroscopic  properties of  brown dwarfs.
However we do not claim  that this CE calculation predicts exactly the
distribution   of   dust   species   in  brown   dwarfs   atmospheres.
Gravitational   settling   effects   would   certainly   change   this
distribution.   We note in  passing that  our main  results \cite[,and
present]{MDpap}  concur  with   both  those  of  \cite{Lodders99}  and
\cite{tsuji73}.   This  consistency between  independent  CE works  is
reassuring.

In    this    work,     contrary    to    recent    calculations    by
\cite{1999ApJ...512..843B},  we  have  not  attempted  to  handle  the
effects  of gravitational settling  (i.e. diffusion  of the  grains to
lower atmospheric layers).  We  do not account for elemental abundance
depletion resulting from dust  grain settling.  This process is likely
important  in the  uppermost layers  of brown  dwarf  atmospheres, and
would tend  to deplete these  layers of their refractory  elements and
dust grains.   However, the error  introduced by this omission  on the
models presented in this paper is small since these represent limiting
cases with non-existent and  complete settling, where the dust opacity
has  been  ignored  altogether  to  recreate  the  latter  case.   The
available abundance of refractory  elements involved in dust formation
can  however be  slightly  overestimated.  The  complete treatment  of
gravitational  settling goes  beyond a  simple parametrization  of the
problem, and involves solving the  diffusion of the dust as a function
of the characteristic timescale of several important processes such as
the sedimentation, coagulation,  condensation and convective mixing of
the  dust.  This  work  is,  however, under  development  and we  will
present our findings in a separate publication.

\begin{figure}[]
\psfig{file=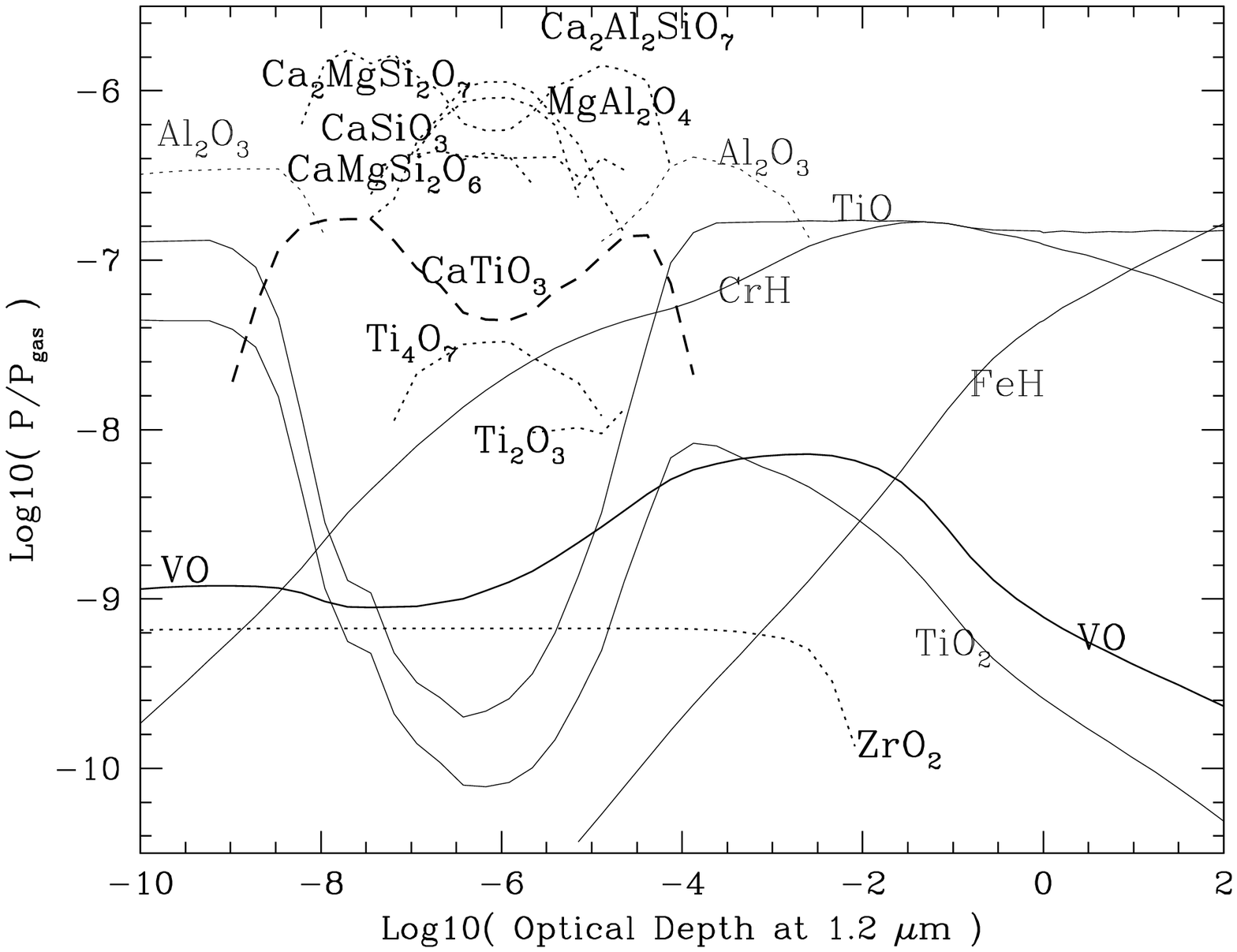,width=\hsize,angle=0}
\caption[]{\label{TiOeos26}  Run  of the  relative  abundances of  gas
phase  (full   lines)  and  crystallized  species   across  a  T$_{\rm
eff}=2600$~K  model atmosphere  typical  of the  young Pleiades  brown
dwarfs Teide1  and Calar3. The condensation  of perovskite (CaTiO$_3$,
dashed  line)  is  the  principle   cause  of  TiO  depletion  in  the
atmospheres  of dwarfs  later than  about  M6.  The  abundance of  the
condensate Ca$_2$SiO$_4$  is drawn at log10($\tau$)$=-5.0$  but is not
labeled for sake of clarity.}
\end{figure}

\begin{figure}[]
\psfig{file=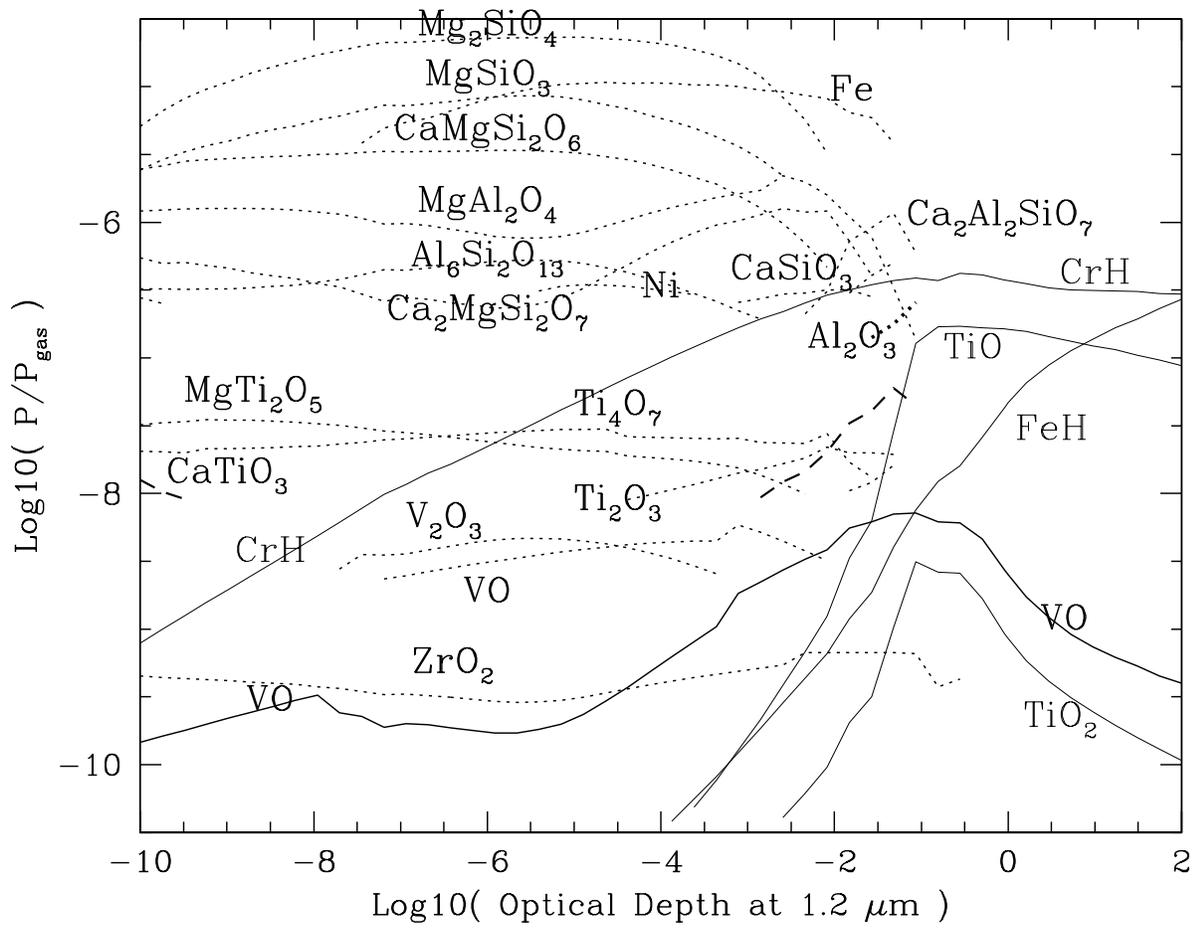,width=\hsize,angle=0}
\caption[]{\label{TiOeos18} Same  as above for  a T$_{\rm eff}=1800$~K
model  atmosphere typical of  the reddest  known field  dwarfs GD165B,
Kelu1, and the DENIS objects.}
\end{figure}

Since the predictions of CE calculations have been described in detail
by \cite{tsuji73}  and \cite{Lodders99}, we  do not deem  necessary to
develop this  here again.  Yet  the completeness of the  grain species
sample included  here (within the  limits of the CE  approximation and
JANAF data) provides an opportunity  to explore some of the effects of
condensation upon  the composition of late-type dwarf  stars and brown
dwarfs atmospheres.  In  the atmospheres of brown dwarfs,  most of the
hydrogen is locked  in H$_2$, most of the oxygen is  in CO, H$_2$O and
SiO, and  most of  the carbon  is in CO  and CH$_4$.   The species
responsible  for   the  strong   optical  to  near-infrared   (0.4  to
1.1~$\mu$m)  opacities in  M dwarf  stars and  young brown  dwarfs are
relatively trace species  much less abundant then CO  or H$_2$O, which
have  large   opacity  cross-sections  per   molecule.   The  relative
abundance of those species are summarized in Figure~\ref{TiOeos26} and
\ref{TiOeos18}  which illustrate  the  nature and  progression of  the
condensation layers into  deeper layers of the photosphere\footnote{In
this paper we  refer to the photosphere as the  part of the atmosphere
which encompasses  the entire optical  depth range where  the spectrum
forms, from the  visible to the infrared ($\approx  \tau = 10^{-4}$ to
1.0).}   as $\teff$  decreases.   At  a $\teff\  =  2600\,$K which  is
typical  of  the  young   Pleiades  brown  dwarfs  Teide1  and  Calar3
\cite[see]{1997ApJ...491L..81Z},  the  clouds  barely  touch  the  top
layers of the photosphere which is located between $\tau_{1.2{\mu}m} =
10^{-4}$ to $1$ depending on  the spectral range considered. As can be
seen from the inner to outer atmospheric regions (right to left on the
plots), the first species to  condense at T$\approx 2000$K is ZrO$_2$,
followed by  corundum (Al$_2$O$_3$)  at T$\approx 1800$K.   We clearly
identify the perovskite  CaTiO$_3$ as the source of  depletion of TiO,
the  principle optical  absorber in  these atmospheres.   However, the
depletion occurs in  this model only above the  photosphere and should
leave  the spectra  relatively  unaffected.  Other  stable species  to
appear  at  T$<1600$K  are  MgAl$_2$O$_4$,  CaSiO$_3$,  Ca$_2$SiO$_4$,
Ca$_2$Al$_2$SiO$_7$, Ca$_2$MgSi$_2$O$_7$, and CaMgSi$_2$O$_6$, as well
as  Ti$_4$O$_7$ and Ti$_2$O$_3$.   These grains  all compete  with the
formation of  CaTiO$_3$ and corundum,  and form an intricate  layer of
clouds just above the photosphere.  Note that, contrary to the reports
by  \cite{tsuji96a}, corundum  is {\sl  not} the  most  abundant grain
species and even disappears in central regions of the clouds.

The    situation   complicates    rapidly   as    $\teff$   decreases.
Figure~\ref{TiOeos18} shows  how the clouds have  already invaded most
of the  photosphere in  brown dwarfs of  about 1800K typical  of field
brown   dwarfs  such   as  GD165B   \cite{1988Natur.336..656B},  Kelu1
\cite{kelu1} or the  DENIS objects \cite{1997A&A...327L..25D}.  Dozens
of  new  grain  species  including  iron,  enstatite  (MgSiO$_3$)  and
forsterite  (Mg$_2$SiO$_4$)  are now  present.   The photospheric  gas
phase abundances  of TiO,  FeH, and CaH  (not shown) are  now strongly
depleted.   This is reflected  by the  gradual disappearance  of these
features  in the  latest-type M~dwarfs  and brown  dwarfs,  a behavior
which is  already apparent  from the observed  spectra of  brown dwarf
candidates BRI0021, GD165B  and Kelu1. The VO abundances  seem, on the
other  hand,  much  less  depleted  by  the  condensation  of  VO  and
V$_2$O$_4$  occurring  only  in  the  upper  photosphere.   And  other
compounds of less-reactive elements such as Li, K, Rb, Cs, and CrH are
left relatively  unaffected, favoring the detection  of their features
in these objects.

Between  $\teff\ =  1800$  and $1000\,$K,  methane (CH$_4$)  gradually
forms at the expense of CO.  The likelihood of detecting methane lines
in the spectra of cool brown dwarfs depends therefore on the height in
the atmosphere  where this transition regime occurs,  and depends upon
$\teff$, gravity,  and dust  opacity conditions such  as gravitational
settling, rotation, winds, etc.

Although the  clouds appear to  persist in Figures  \ref{TiOeos26} and
\ref{TiOeos18} out to the outer edge of the photosphere in our coolest
models,  this only  reflects  the omission,  mentioned  above, of  the
gravitational settling of the grains.  Clouds form more likely in thin
decks above  the deepest condensation layer in  brown dwarfs \cite[see
e.g.]{tsuji99,Ack00}.

The principle  impact of condensation on the  photospheres and spectra
of cool  dwarf is  a gradual depletion  of their  refractory elements,
especially zirconium, titanium, silicon, calcium, magnesium, aluminum,
iron  and nickel.   Clearly,  it is  crucial  for the  balance of  the
opacities in  the models to account  for the leading  grain species in
the chemical equilibrium.  Partially accounting for condensation leads
to errors of  several orders of magnitudes in  the model opacities and
predicted  fluxes.  See  \cite{1998bdep.conf..370A}  for a  comparison
between existing dusty models.   To fully understand the spectroscopic
and photometric properties of brown dwarfs, one must also consider the
optical and radiative properties of dust grains.

\section{\label{clouds}Dust Clouds Construction and Opacities}

In this  paper, we account for  the dust opacity only  in the limiting
case  of  the  AMES-Dusty   models  where  gravitational  settling  is
neglected and which  are relevant for the analysis  of hot red dwarfs.
In  these models, cloud  layers build  up automatically  following the
condensation  equilibrium  which  determines  the  atmospheric  layers
occupied by the grains.  The study of the jovian planets suggests that
cloud  layers are  not  generally distributed  homogeneously over  the
atmospheric surface.  But if the  nucleation process of dust grains is
favored  by  a   combination  of  high  gas  densities   and  low  gas
temperatures,  hotter red  dwarfs  could be  expected  to retain  more
easily  smaller  grains  in  their  photosphere, and  have  them  more
uniformly  distributed over  the stellar  surface.  In  this  work, we
account for the average effect of  the presence of clouds on the model
structures and emitted spectra.   We therefore assume a plane-parallel
symmetry,  i.e.   an homogeneous  distribution  of  clouds across  the
surface of  the brown  dwarfs.  The spectral  distribution of  a brown
dwarf with  a more complex ring  pattern of clouds  familiar to jovian
planets  could, in  the end,  be reconstructed  from a  mosaic  of the
present  models until  a  full three  dimensional calculation  becomes
possible.

Early  attempts  to compute  the  opacities  of  grains were  made  by
\cite{Cameron&Pine73} and  \cite{Alex75}.  More detailed calculations,
including the  effects of chemical equilibrium  calculations and grain
size    distributions    were    reported   by    \cite{Alex83}    and
\cite{Pollack85}.  \cite{Alex94} have described the computation of the
opacity  of  grains with  the  inclusion  of equilibrium  condensation
abundances, the  effects of the  distribution of grain sizes,  and the
effect of grain shape through the Continuous Distribution of Ellipsoid
(CDE) model  of \cite{Bohren&Huffman83}.  These  calculations included
the  absorption  and  scattering  due to  magnesium  silicates,  iron,
carbon,  and silicon  carbide  grains  for a  wide  range of  chemical
compositions down to 700~K.  We  have explored the CDE method employed
by Alexander \&  Ferguson, but have retained a  purely spherical shape
of the grains in the present study for simplicity.  Another difference
with  Alexander \&  Ferguson  is that,  instead  of approximating  the
number density of grains indirectly, these quantities are now provided
by our chemical equilibrium as described in Section \ref{eos}.  26 new
condensates have  been added to  the original list  of Fe, C,  SiC and
magnesium  silicates, for  a total  of 30  among which  are MgSiO$_3$,
Mg$_2$SiO$_4$,  Al$_2$O$_3$  and  MgAl$_2$O$_7$, using  polarizability
constants   from    laboratory   studies   by   \cite{Tropf&Thomas90},
\cite{Loike95},     \cite{Begemann97},     and     \cite{Dorschner94}.
Figure~\ref{TiOeos26}  and  \ref{TiOeos18}  suggest that  the  calcium
silicates  can also play  an important  role in  the opacity  of brown
dwarf   atmospheres.   In  fact,   complex  calcium   silicates  (here
Ca$_2$Al$_2$SiO$_7$,  Ca$_2$MgSi$_2$O$_7$,  and  CaMgSi$_2$O$_6$)  are
amoung the most abundant species  in the layers where these grains are
present.  Since, for the latter two species, no data were available to
construct  their opacity  profiles,  we have  simulated their  opacity
using  the profile  of Ca$_2$Al$_2$SiO$_7$.   We includes  in general,
more  accurate   number  densities,  and  better   and  more  complete
cross-sections of dust grains than included in \cite{Alex94}.  Updated
Rosseland and  Planck opacities computed with  these updated opacities
will be published in more details separately \cite[]{Alex00}.

\begin{figure}[]
\psfig{file=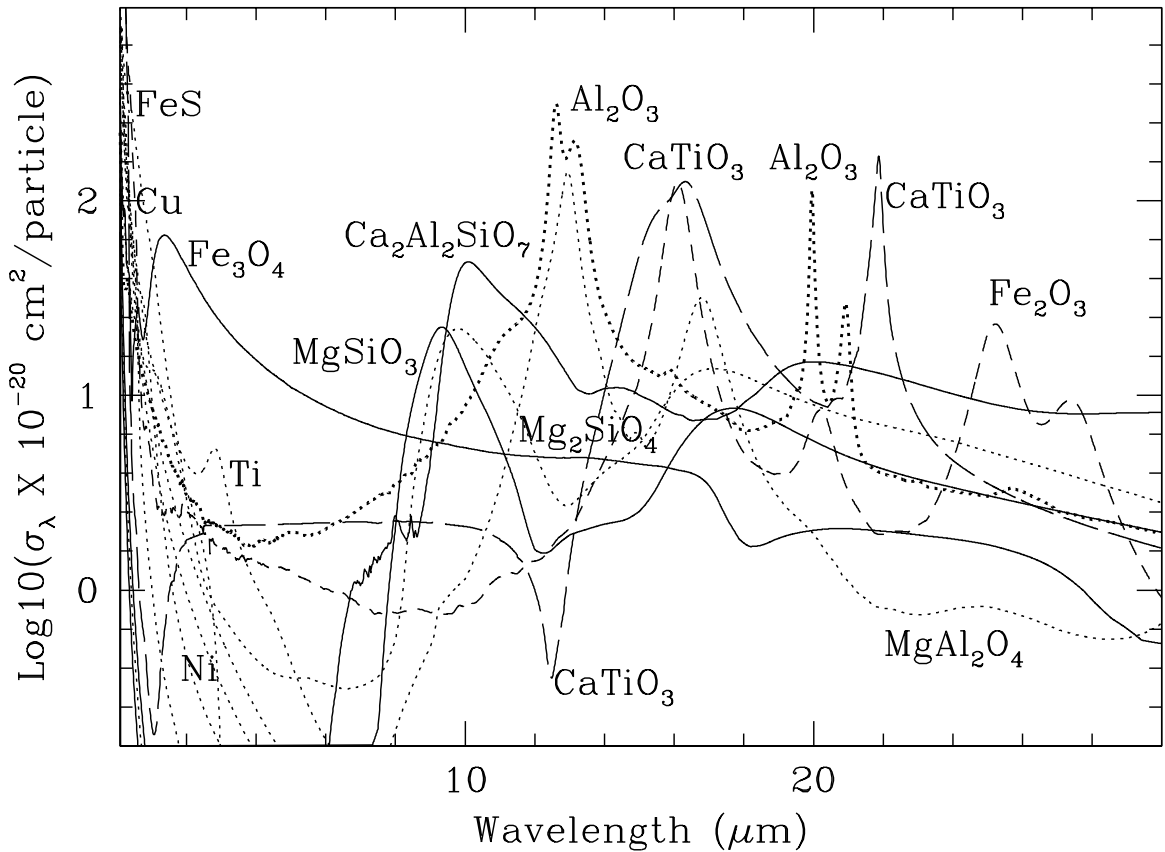,width=\hsize,angle=0}
\caption[]{\label{opacIR}  The extinction cross-sections  per particle
of  dust grains.   The  mie  formalism is  used  assuming a  power-law
($\alpha=-3.5$)  grain size distribution  with diameters  from 0.00625
and 0.24 $\mu$m.  Monoatomic grains  such as Fe, Cu, and Ni contribute
scattering  at  optical wavelengths  only,  while corundum,  magnesium
aluminium   spinel,  calcium   titenide,   hematite,  magnetite,   and
Ca$_2$Al$_2$SiO$_7$  crystals  show  strong  peaks  of  absortion,  at
infrared wavelengths,  that could compete  with the local  water vapor
``continuum''  in hot/young brown  dwarfs.  Note  however that  if the
grains  were elliptical  and randomly  oriented, the  sharp absorption
peaks shown here could be washed out. }
\end{figure}

The opacity profiles of these grains are shown in Figure~\ref{opacIR}.
Most  spectral distributions  seen in  this plot  are  pure absorption
profiles.  Scattering  contributes only  at UV to  optical wavelengths
for  the  grain   sizes  adopted.   Corundum,  enstatite,  forsterite,
hematite,   magnetite,   and   Ca$_2$Al$_2$SiO$_7$   have   absorption
cross-sections  exceeding those  of  water vapor.   In  a brown  dwarf
atmosphere, however, water  is at least two orders  of magnitudes more
abundant then most of these grain species, so that water vapor remains
the  leading opacity  source  between 1  and  8 $\mu$,  and beyond  20
$\mu$m.  In other words, grains do not contribute significantly to the
opacities in  the near-infrared where  water bands still  dominate the
brown  dwarf spectra  from  $J$ to  $K$  (1.0 to  3~$\mu$m).  We  have
demonstrated   in   Section~\ref{eos}    that   grain   species   have
concentrations  similar to  those of  TiO, VO  and most  other optical
absorbers.   The  impact of  the  grain  opacities  is, therefore,  to
enhance and gradually replace the optical opacities as these gas phase
species disappear via condensation.

\begin{figure}[]
\psfig{file=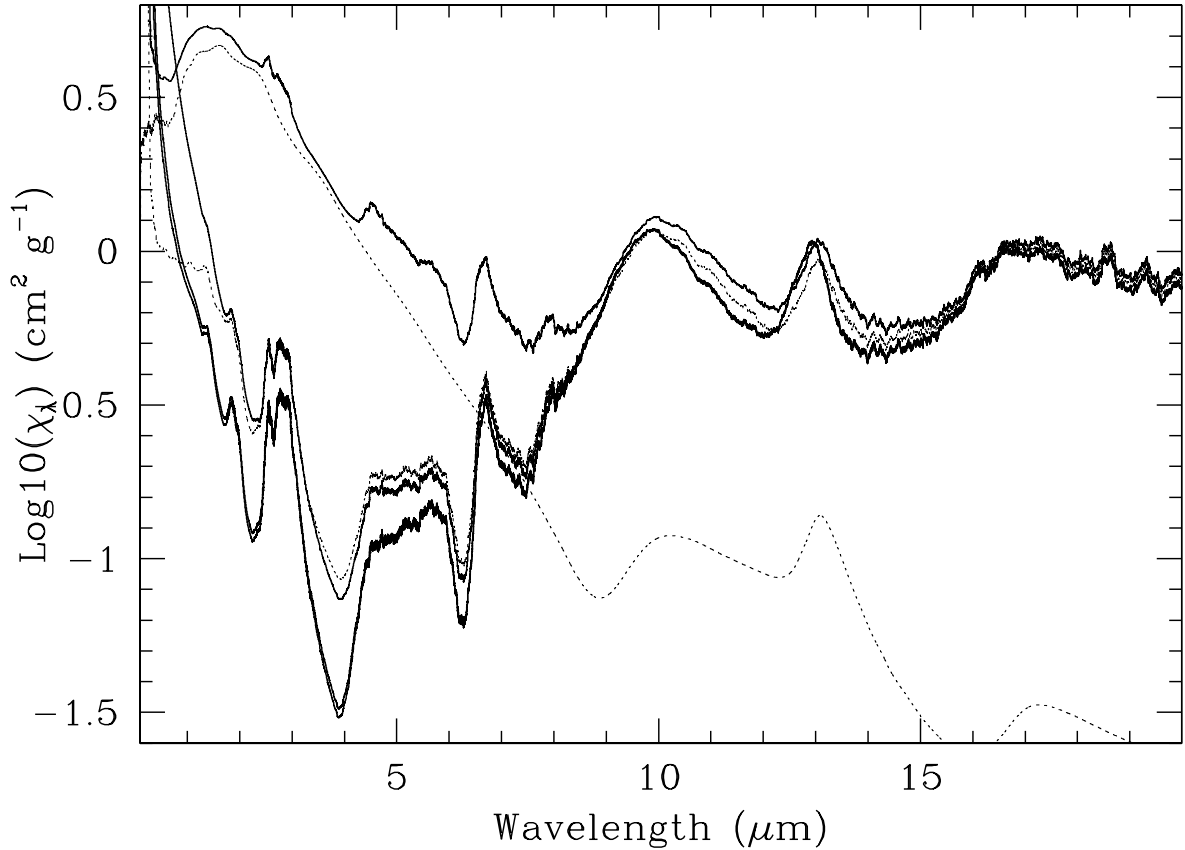,width=\hsize,angle=0}
\caption[]{\label{gsize}  The  extinction  profiles are  compared  for
grain size  distributions with 1, 2,  10 and 100 times  the ISM values
adopted for  this work  (full lines from  bottom to  top respectively,
where  the  two  first  curves  are  nearly  undistinguishable).   The
scattering  and absorption contributions  of the  100 ISM  profile are
also  shown   (dotted  lines).   The  conditions  are   those  of  the
photospheric    layers   ($\tau_{1.2{\mu}m}\approx    10^{-4}$,   i.e.
T$\approx 1300$K)  of our standard 1800K  AMES-Dusty model atmosphere.
The structures seen  in the profile at $\lambda >  8.5\mu$m are due to
dust absorption (Mg$_2$SiO$_4$ at 10 and 16.5 $\mu$m and MgAl$_2$O$_4$
at 13 $\mu$m).  Scattering contributions dominate below $0.5\mu$m, and
remain modest  at longer wavelengths  for grain sizes $\leq  10$ times
the  adopted values.   The absoption  profile, on  the other  hand, is
only little sensitive to grain sizes. }
\end{figure}

The extinction caused  by grains in a stellar  atmosphere also depends
on the  rate of grain  formation and the resulting  size distribution.
For all grains included we  have assumed, as in Alexander \& Ferguson,
an interstellar size distribution of the grains with diameters ranging
from  0.00625  to 0.24~$\mu$m.   For  comparison, \cite{tsuji96a}  and
\cite{tsuji96b} assumed grains with  a fixed diameter of 0.1~$\mu$m in
their  model  atmosphere  calculations.   Although those  choices  are
purely  arbitrary,  the  consequences  are  minimal  since  the  grain
diameter  cancels out  in the  opacity  calculations as  long as:  (1)
abundance conservation  is assumed i.e.  larger grains  must lock more
particles, reducing the  number of grains per gram  of stellar plasma,
and, (2)  that the cross-sections  behave in the Rayleigh  limit, i.e.
the  wavelength is larger  than the  size of  the grains.   Our tests,
shown in Figure~\ref{gsize}, confirm that this is the case for 1 to 10
$\mu$m-size  grains,  but  they  also  indicate  that  the  scattering
increases  rapidly for  sizes larger  than  10 $\mu$m,  i.e. when  the
Rayleigh limit beaks down over  the wavelengths carrying flux in these
objects.  But even  is the opacities are sensitive  to the grain sizes
beyond the  Rayleigh regime, the enormous scattering  effects seen for
100 $\mu$m grains in  Figure~\ref{gsize} seems to exclude the presence
of such grains in brown  dwarfs.  We therefore believe that such large
grains tend rapidly  to become larger by coagulation  to be eliminated
by  sedimentation  in  these  high-gravity atmospheres.   An  accurate
answer to this question can only come from time-dependent grain growth
calculations  incorporating the  effects of  sedimentation, diffusion,
coagulation  and coalescence  for the  conditions prevailing  in brown
dwarfs atmospheres.   See \cite{Alex00} for a  detailed description of
the dust opacities used in the present models.

\section{\label{molec}Molecular Opacities}

Molecules dominate  the spectral distribution of  brown dwarfs.  Young
brown  dwarfs   hotter  then  2000K  emit  flux   principally  in  the
near-infrared  ZJHK bandpasses from  0.9 to  2.5~$\mu$m with  bands of
water  steam in  the infrared  and  TiO and  VO bands  in the  optical
shaping each side of this spectral distribution.  Below 2000K, TiO, VO
and  CaH  bands  vanish as  a  result  of  the condensation  of  these
refractory elements  and the collision-induced opacities  of H$_2$ and
the growth of  CH$_4$ bands become increasingly important  in the H, K
and L  bandpasses at 1.6, 2.0,  and 3.5~$\mu$m.  It  is this situation
that causes the radiation to  remain forced to emerge principally from
the near-infrared  bandpasses rather  than emerging redwards  as black
bodies would  do.  Several other  molecules are also present  in brown
dwarf  spectra; overtones  of  CO  at 2.3  and  4.7~$\mu$m are  strong
T$_{\rm eff}$ indicators  and play an important cooling  role upon the
upper  atmospheric structure,  vibrational bands  of SiO  and hydrides
such  as  OH,  SiH  and  MgH  determine  the  ultraviolet  and  visual
radiation, the  Wing-Ford system of FeH  at 0.98~$\mu$m is  one of the
most prominent  features in the  spectra of late-type stars  and young
and/or massive brown dwarfs.

Our  molecular opacity data  base includes  (i) a  list of  43 million
atomic transitions by \cite{cdrom1}, (ii) Collision-Induced Absorption
(CIA) opacities  for H$_2$,  He, H, N$_2$,  Ar, CH$_4$, and  CO$_2$ by
\cite[][and   references  therein]{h2h2a,h2hea,n2n2a,n2n2b,h2n2,n2ch4,
ch4ar,h2ch4,ch4ch4,co2co2}, (iii) ab initio  line lists for H$_2$O and
TiO by  \cite{ames-water-new,ames-tio}, (iv) CO from the  line list by
\cite{Goorvitch&Chackerian94a,               Goorvitch&Chackerian94b},
\cite{Goorvitch94}, (v)  most other molecular systems such  as MgH and
OH are included from the list  by \cite{cdrom15}, (v) VO and CrH lines
have been calculated by R. D.   Freedman for this work, while (vi) FeH
lines from \cite{FeHberk2} have been also included.  For the remaining
molecular band  systems for which no  line lists were  available to us
(CaH)  we apply  the  Just Overlapping  Line  Approximation (JOLA)  as
described  and utilized  by  \cite[][and references  therein]{tsuji95,
tsuji96a,tsuji96b}.

We  have  also  included  a  combination of  the  \hitran\  and  GEISA
databases \cite[]{hitran92,geisa92} summing  up to about 700,000 lines
of 31 molecules  with a total of 74 isotopes.   The molecules with the
largest number  of lines  which are included  in our models  are O$_3$
(168,881 lines), CO$_2$ (60,790),  and CH$_4$ (47,415).  This database
only includes  the strongest lines  of these molecules.   However, the
gf-values and  position of the lines have  comparatively high accuracy
and allow us to diagnose  their importance in brown dwarf atmospheres.
A comparison  of models computed  with these opacities to  the NextGen
models \cite{ng-hot,araa} can be found in \cite{AMESpap}.

\section{\label{model}The Model Atmosphere}

We use the  model atmosphere code \phx\ (version  10.8).  The original
versions  of  \phx\ were  developed  for  the  modeling of  novae  and
supernovae ejecta described  by \cite[][and references therein]{jcam},
and  is dotted of  a detailed  radiative transfer  \cite[]{s3pap} that
allows for  spherical symmetry.  Its  more recent application  to cool
dwarfs is  described in detail by \cite{MDpap,ng-hot},  and has served
to  generate grids  of  stellar model  atmospheres which  successfully
described     low     mass      stars     in     globular     clusters
\cite[]{BCAH95,clusterpap}  and   the  galactic  disk   main  sequence
\cite[]{BCAH98}.  These  former model grids  are known to  the stellar
community as the  1995 Extended and the 1996-1999  NextGen models, and
allowed  a preliminary  incursion  into the  regime  of evolved  brown
dwarfs down to  T$_{\rm eff} = 1600\,$K (Extended  models) and to 900K
\cite[]{gl229b}.   These  models  successfully predicted  the  general
spectroscopic properties  of evolved brown  dwarfs {\sl prior}  to the
discovery  of Gliese  229B, which  then helped  confirm its  very cool
brown  dwarf nature  \cite[]{1995Natur.378..441A}.  Yet  the  lack, in
these essentially stellar models, of dust condensation in the chemical
equilibrium made them inadequate to model in detail such cool objects.

The addition to \phx\ of the treatment of condensation in the chemical
equilibrium, and  of dust clouds,  as described in  Sections \ref{eos}
and \ref{clouds},  was completed in  1996 \cite[]{1998bdep.conf..370A,
1998bdep.conf..438A,1998csss...10...63A},   and   served  to   compute
M~dwarfs  and brown  dwarfs model  atmospheres, synthetic  spectra and
broadband   colors  for   specific  analysis   \cite[]{legg96,  kelu1,
tinney98,  legg98,  Martin98,  gd165b,  lhs1070,  legg00,basri00}  and
interior models  \cite[]{CBAH00}.  In this paper we  present the final
version  of these  models in  two limiting  cases:  (1) ``AMES-Dusty''
which include both the dust  formation in the chemical equilibrium and
opacities,  and;  (2)  ``AMES-Cond''  which  include  the  effects  of
condensation in  the chemical equilibrium  but ignores the  effects of
dust opacities  altogether.  This latter  case is computed  to explore
the  case  where  dust   grains  have  formed,  but  have  disappeared
completely (eg by sedimentation  i.e. settling below the photosphere).
These  two model sets  also distinguish  themselves from  the standard
NextGen models by  the use of the NASA AMES H$_2$O  and TiO line lists
while    the    NextGen     models    were    computed    using    the
\cite{TiOJorg,schryb94} line  list.  This  choice is motivated  by the
incompleteness of the 1994 lists  to high gas temperatures (T$_{gas} >
2000$K) as discussed in \cite{AMESpap}.

For the  purpose of  this analysis, we  use the radiative  transfer in
plane-parallel mode.   The convective  mixing is treated  according to
the  Mixing Length  Technique (MLT).   We  consider pressure-dependent
line-by-line opacity sampling treatment  for both atomic and molecular
lines in  all models.  We  do {\sl not} pre-tabulate  or re-manipulate
the  opacities   in  any  way:  \phx\   includes  typically  $\approx\
15\alog{6}$ molecular and atomic  transitions which are re-selected at
each model  iteration and each  atmospheric depth point from  our data
base   described   above.   The   lines   are   selected  from   three
representative  layers of the  atmosphere at  each model  iteration to
ensure  consistency  of  the  calculation.   Van  der  Waals  pressure
broadening of the  atomic and molecular lines is  applied as described
by  \cite{vb10pap}.  We neglect  the effects  of convective  motion on
line formation  since the velocities  of the convection cells  are too
small  to  be detected  in  low-resolution  spectra  and will  have  a
negligible influence on the transfer of line radiation.

A trial  atmospheric profile is applied, the  equations of hydrostatic
and radiative  transfer are solved,  and the solution is  tested until
convergence is  reached.  The model  is considered converged  when the
energy is conserved  within a tenth of a percent  from layer to layer.
At each of the model iterations, a spectrum with typically over 30,000
points is  generated which samples  the bolometric flux from  0.001 to
500~$\mu$m with a step of 2{\AA}  in the region where most of the flux
is  emitted  (i.e.   0.1  to  10~$\mu$m).   The  final  spectrum  must
generally  be {\sl  degraded}  to the  instrumental resolution  before
being  compared  to low-resolution  observations  of  stars and  brown
dwarfs.   The model  atmospheres  are characterized  by the  following
parameters:  (i) the  surface gravity,  $\log(g)$, (ii)  the effective
temperature, $\teff$\unskip,  (iii) the mixing length  to scale height
ratio,  $\alpha$, here  taken to  be unity,  (iv)  the micro-turbulent
velocity $\xi$,  here set to  $2\kms$, and (v) the  element abundances
taken from \cite{abund}.

For this paper  we have calculated a uniform  grid of AMES-Cond models
ranging from $\teff = 3000$ to  100K in 100K steps, and with gravities
ranging  from $\log  g=  2.5$ to  6.0 in  steps  of 0.5  dex at  solar
metallicity.  The  AMES-Dusty grid was calculated from  $\teff = 3000$
to 1400K in steps of 100K,  with gravity ranging from $\log g= 3.5$ to
6.0 in steps of 0.5 dex.  All models were fully converged.

Although \phx\ can  treat the effects of external  radiation fields on
the model  atmosphere and the  synthetic spectrum \cite[]{Baron93},
we assume  here a negligible external radiation  field for simplicity.
It is clear, however, that  UV radiation impinging on the brown dwarf,
from a hotter  companion, will change the structure  of the atmosphere
and the corresponding spectra.   We are investigating these effects in
a separate publication \cite[]{Barman00}.

\section{\label{struc}Atmospheric Structures and Convection}

\begin{figure}[]
\psfig{file=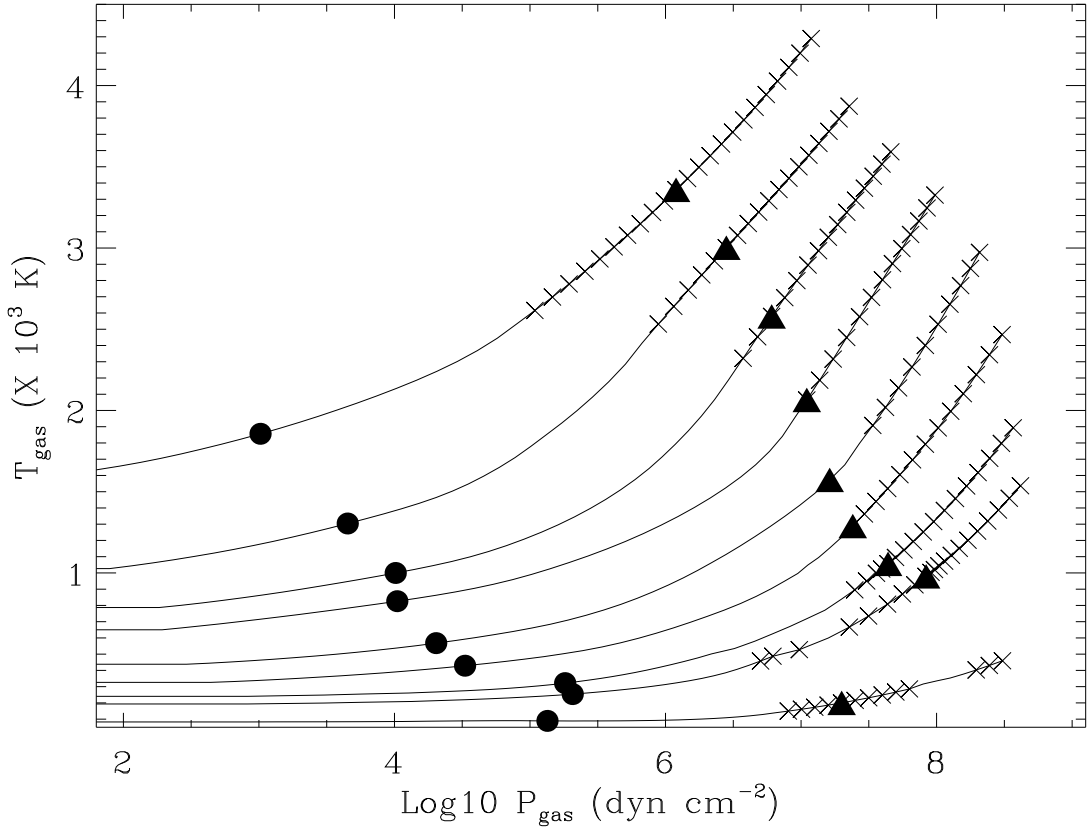,width=\hsize,angle=90}
\caption[]{\label{condstruc}Thermal structures of the AMES-Cond models
with T$_{\rm  eff}$ ranging from 3000  to 100K by steps  of 500K, with
two  additional   models  at  700   and  400K,  logg=5.0,   and  solar
metallicity.   The convection  zones are  labeled  with cross-symbols.
The approximate  location of the photosphere is  indicated with filled
circles and triangles marking  the $\tau_{\rm 1.2{\mu}m}= 10^{-4}$ and
1.0 optical  depths.  All models shown stop  at $\tau_{\rm 1.2{\mu}m}=
10$.  As $\teff$ decreases, the photosphere becomes progressively more
isothermal.}
\end{figure}

The  photospheric  thermal structures  of  the  AMES-Cond models  with
T$_{\rm   eff}$  ranging   from  3000   to  100K   are   displayed  in
Figure~\ref{condstruc}.   The   convection  zones  are   labeled  with
cross-symbols.    As  $\teff$   decreases,  the   photosphere  becomes
progressively more isothermal.  While  the convection zone retreats to
deeper layers down to $\teff = 1000$K, an outer convection zone begins
to form in  the clouds until this zone detaches  itself from the inner
convection regime  in models cooler  than 500K.  Meanwhile,  the inner
convection  continue  to retreat  inwards.   This  appears to  confirm
qualitatively     earlier     work     by     \cite{Guillot94}     and
\cite[]{1997ApJ...491..856B}.   Yet even  in our  coolest  models, the
inner convection zone  always reaches at least up  to an optical depth
of $\tau_{1.2  {\mu}m} =  10$.  For $\teff  = 1000$K for  example, the
convection  zone seems  to be  quite deeper  in Burrows  \etal\ (1997)
models (roughly $P_{gas} > 100$ bar  as seen from their Figure 5) than
in our  models.  Note  that we treat  the convection according  to the
Mixing  Length Theory  from the  onset of  the  Schwarzchild criterion
while Burrows \etal\ (1997) assumed a pure adiabatic mixing throughout
the  convective  unstable zones.   But  this  appears  to be  a  valid
approximation   since  our   calculations  indicate   that   the  true
temperature gradient as predicted by  the MLT remains within 0.05\% of
the adiabatic gradient value at each layer.  So the difference appears
to lie in the opacities included in the construction of the respective
models: their models would be more transparent to radiation than ours.

The  optical  resonance  lines  of  K~I  and  Na~I~D  also  contribute
significantly  to   the  optical  opacity  and  the   heating  of  the
atmospheric  layers.  We  have explored  their impact  on  the thermal
structure  of a  $\teff =  1000$K,  $\log g=  5.0$, solar  composition
model.  It  appears that their  opacity contribution accounts  for 100
and 300K of heating in the  photospheric ($\log P = 5.5$) and internal
($\log P =  8.5$) layers respectively.  The models  become unstable to
convection further out when atomic lines are included ($\log P = 7.76$
versus  $8.1$).  And  uncertainties  in the  applicability of  Lorentz
profiles (estimated  from models computed with  restricted coverage of
the   line  wing  opacity   contributions)  produce   a  corresponding
uncertainty  of less  than  40K in  the  photosphere and  150K in  the
internal  layers.    These  uncertainties  are   therefore  of  little
importance for the synthetic spectra and evolution models, compared to
those  tied  to  the  treatment  of  the dust  (Cond  vs  Dusty),  and
incomplete  molecular opacities  (e.g.  H$_2$O  opacity  profile.  See
Allard,  Hauschildt   \&  Schwenke,  2000\nocite{AMESpap}).   However,
neglecting  the K~I  and Na~I~D  doublet opacities  altogether  in the
construction of the thermal structures  has a greater impact and fully
explains the difference between our models and those of Burrows \etal\
(1997).  Indeed, while our model at  $\teff = 500$K and $\log g = 5.0$
do  not present detached  convection zones,  we reproduce  exactly the
several  detached   convection  zone  found  by   these  authors  when
neglecting  atomic line opacity  in the  model construction.   We must
conclude from this that these  opacities were neglected in their work.
The reality of the occurrence of detached convection zone is therefore
likely closer to our predictions.

\begin{figure}[]
\psfig{file=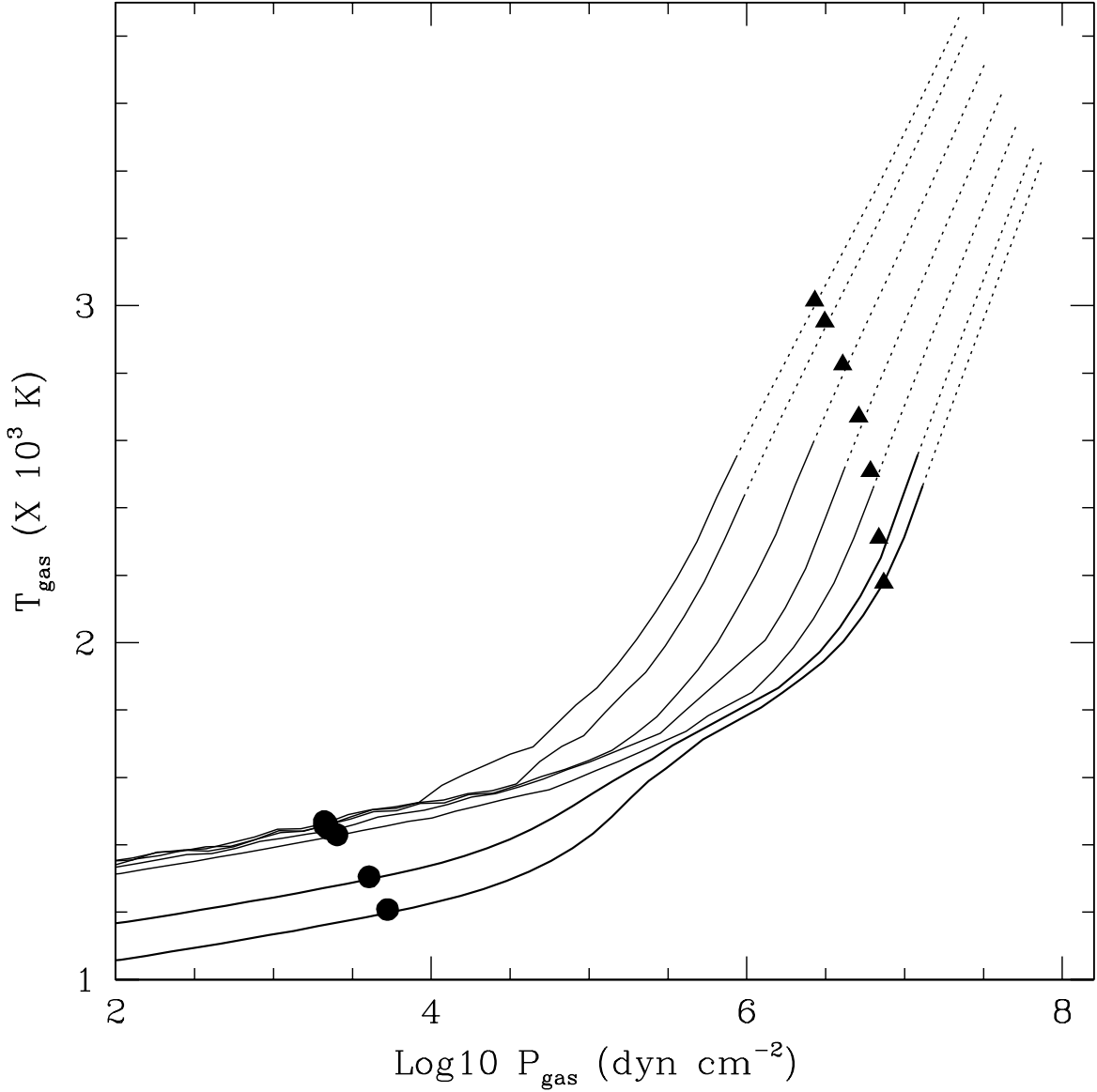,width=\hsize,angle=0}
\caption[]{\label{conv} Thermal  structures of the  fully dusty models
with T$_{\rm eff}$  ranging from 2400 to 1600K by  steps of 200K, with
two additional models at 2500  and 1500K, logg=5.0. None of the curves
shown  actually overcross.   The radiative  zones are  marked  by full
lines  while the  convective region  is  shown as  dotted lines.   The
location of the  photosphere is also indicated, with  full circles and
triangles marking the $\tau_{\rm  1.2{\mu}m}= 10^{-4}$ and 1.0 optical
depths  respectively.   The  strongest  optical  molecular  bands  and
resonance lines form near $\tau_{\rm 1.2{\mu}m}= 10^{-4}$.  All models
shown stop at $\tau_{\rm 1.2{\mu}m}= 10$.}
\end{figure}

The thermal structures  of the fully dusty AMES-Dusty  models over the
T$_{\rm  eff}$-range where  dust begins  to form  (2500 to  1500K) are
displayed  in Figure~\ref{conv} at  constant gravity.   The convection
zone, marked by dotted lines, extends  out to T$_{gas} = 2500$K is all
these  models.  This corresponds  to optically  thin layers  in models
hotter than 1600K.   Even down to 500K, these  dusty atmospheres never
become fully radiative.   But the interesting part is  what happens to
photospheric regions  as grain  opacities begin to  heat up  the outer
layers.   Within the  photosphere  (marked by  with  full circles  and
triangles), the temperature normally decrease smoothly with decreasing
$\teff$,  and the  thermal structures  parallel for  grainless models.
Here, the greenhouse effect of the dust tends to raise the temperature
of the  outer layers increasingly  with decreasing $\teff$.   This has
for effect that the outer  structures level off between $\teff = 2600$
and  1800K to a  T$_{gas}$-value in  a narrow  range between  1280 and
1350K.  The slope of the  thermal structure in the line forming region
becomes therefore  increasingly flatter in  that $\teff$-range.  Below
1800K, the greenhouse effect saturates and the outer thermal structure
resumes its  decrease in temperature  with decreasing $\teff$.   It is
interesting to note that 1800K  is also the break-up temperature where
full-dusty  atmospheres become unrealistic  in modeling  brown dwarfs.
This  can be  seen from  Figure 6  of \cite{CBAH00}  and  from Section
\ref{color} below.  We believe that grains sedimentation has certainly
started    at    these    temperatures    as   also    concluded    by
\cite{lunine89}, \cite{1999ApJ...512..843B} and \cite{Ack00}.

\section{\label{spec}Synthetic Spectra}

In Figure  \ref{Tseq_cond} we display  the spectral sequence  of brown
dwarfs to  extrasolar giant planets (hereafter  EGP) model atmospheres
from  $\teff =  3000$  to  200K in  the  total gravitational  settling
(AMES-Cond) approximation.   All dust  opacity is neglected,  but also
all   optical  molecular   opacity  sources   disappear  due   to  the
condensation of species involving Ti, V, Ca and Fe (TiO, VO, CaH, FeH,
etc.),  making  these models  transparent  to  the emergent  radiation
bluewards of 1.0 $\mu$m.  Because  of the absence of dust opacity, the
photospheric   layers   are  very   cool   compared  to   non-depleted
atmospheres.   The formation  of  optical atomic  resonance lines  and
infrared molecular bands is then favored.  We observe that water vapor
bands (0.93, 0.95,  1.2, 1.4, 1.8, 2.5, and  5-10~$\mu$m in the window
shown by  this plot) increase  rapidly in strength.   Another striking
consequence of the cool  photospheric temperatures is the formation of
CH$_4$  bands (3.5  and  6-10~$\mu$m,  with weaker  bands  at 1.6  and
2.2~$\mu$m  appearing in  cooler  models) already  at 2000K.   Methane
gradually replaces water vapor bands while H$_2$O condenses out to ice
below 300K.

One major feature of the  AMES-Cond model spectra is the extraordinary
growth  of atomic  resonance  absorption lines  at short  wavelengths.
\cite{BMS00}  have explored  grainless  models of  methane dwarfs  and
found that van der Waals  broadening of K~I and Na~I resonance optical
lines can extend to several thousands of Angstroms on each side of the
line cores.  Our models  behave similarly.  Figure \ref{vdw} shows how
the  van der  Waals wings  of the  Na~I~D and  K~I resonance  lines at
$\lambda$5891,5897\AA\  and $\lambda$7687,7701\AA\  completely depress
the optical flux  of cool brown dwarfs.  In  this case ($\teff=1000$K,
$\log g= 5.0$), the wings extend largely over 7000\AA\ on each side of
the line  center.  This is as  large as hydrogen Balmer  line wings in
cool white dwarfs!  However, to our knowledge, it is the first case of
such behavior in metal  lines encountered in stellar astronomy.  While
the van der Waals collisional  C6 damping constant may be sufficiently
accurate for the  treatment of alkali element lines  in the hydrogenic
approximation  in low  mass stars  and  red dwarfs  where these  lines
rarely exceed a width  of 50{\AA}, this treatment becomes questionable
under these unprecedent conditions  as also concluded by \cite{BMS00}.
Here, we observe  for example that the red  wings of these transitions
prevents  even a  fraction of  the flux  from escaping  in  the J-band
window  around 1.25~$\mu$m, while  observed spectral  distributions of
methane dwarfs tend to carry more flux in this window.  The reason for
such  large line  broadening  is  not the  decreasing  $\teff$ of  the
photospheric  gas  pressure.   It  is  rather,  as  also  observed  by
\cite{allard90,MDpap,tinney98}  for  metal-depleted  atmospheres,  the
result of  the increasing transparency of the  atmosphere which allows
us to  see deeper  into the structure  to inner high  pressure depths.
The  line wing flux  integrates therefore  over an  increasingly large
column density of the atmosphere as optical molecular opacities vanish
via condensation.

Figures  \ref{Tseq_cond_optH}  and  \ref{Tseq_cond_optC}  display  the
change of  the optical  to red spectra  as a function  of temperature,
where the  gradual disappearance  of TiO, VO,  FeH, and CaH  bands (by
condensation of related species)  and gradual strengthening of optical
Na~I~D  and   K~I  lines  becomes  obvious.   The   TiO  band  systems
($\lambda$0.545,  0.616,  0.639,  0.665, 0.757,  0.774,  0.886~$\mu$m)
become     undetectable     below     2000K,     while     the     MgH
($\lambda$0.513~$\mu$m),  CaH  ($\lambda$0.694  and 0.706~$\mu$m),  VO
($\lambda$0.829, 0.848  and 0.961~$\mu$m), FeH ($\lambda$0.990~$\mu$m)
bands persist down to  1500K.  The CrH bands at $\lambda$0.861~$\mu$m,
already visible at  2500K in these AMES-Cond models,  grow in strength
as  $\teff$   decreases  until   it  disappears  by   condensation  of
Cr$_2$O$_3$ below  900K.  However, we  must point out that  red dwarfs
are  heavily reddened  by dust  opacities at  least down  to  $\teff =
2000$K, so that the Cond  models overpredict the strength of CrH bands
over that temperature range.  Still,  clearly the CrH bands become one
of the strongest molecular system to be observed in the red spectra of
cooler brown dwarfs.

From   2000K,    we   also   see    the   H$_2$O   band    system   at
$\lambda$0.927~$\mu$m  becoming  increasingly  stronger.   The  Na~I~D
resonance doublet remains visible down to 400K, and K~I already begins
to get  locked into  dust below about  900K, a temperature  typical of
currently known methane dwarfs such as Gl229B.  Other features growing
in strength as $\teff$ decreases are the lines of alkali elements such
as Li~I  at $\lambda$6708\AA, Rb~I at $\lambda$7802  and 7949\AA, Cs~I
at $\lambda$8523  and 8946\AA, and Na~I at  $\lambda$8185\AA.  We also
note  the presence of  diagnostic lines  further in  the near-infrared
such    as    the   K~I    doublet    at   $\lambda$11693,11776    and
$\lambda$12436,12525\AA,and a Na~I line at $\lambda$11409\AA.

The  $\lambda$6708\AA\  Li  line,   normally  used  to  determine  the
substellar   nature  of   brown   dwarfs  \cite[]{Rebolo92},   remains
detectable  down  to  700K.   The  Na~I.  Rb~I  and  Cs~I  lines  keep
increasing  in  strength,  but  this  is likely  an  artifact  of  the
inevitable   incompleteness  of   thermochemical   databases  in   the
construction of the chemical composition at these temperatures.

Thanks to  R. Freedman (NASA-Ames), we  were able to  replace the band
model approximation by a detailed  line list for VO besides also being
able to include  CrH lines for which we  had no previous counterparts.
The result is  that the present models show  weaker VO bands, relative
to TiO,  strength than in  previous models.  A detailed  comparison to
high resolution observations of M  and brown dwarfs is being published
separately \cite[]{Schweitzer00}.  However, note that the VO line list
does  not include  C-X  system at  0.75~$\mu$m.   Our current  models,
therefore, overestimate the flux in the 0.75~$\mu$m region.

In  Figure   \ref{Tseq_cond_co},  we  explore  the   behavior  of  the
4.55~$\mu$m CH$_3$D band  system between $\teff = 2000$  and 400K.  At
those  wavelengths, H$_2$O  provides the  pseudo-continuum absorption.
In the limit  of the Cond models, CH$_3$D  is practically undetectable
until it  begins to grow  from 1000K to lower  effective temperatures.
The ammonia band at around 11  $\mu$m behaves similarly as can be seen
from  Figure \ref{Tseq_cond_nh3}.   This is  a result  of  the growing
transparency of the  atmosphere while water begins to  condense in the
uppermost layers of these Cond models.  Note that although CO bands at
4.67~$\mu$m are  not visible  in Cond models  with $\teff  \ge 1800$K,
these bands do  appear in corresponding brown dwarfs  and stars.  This
is because the Cond limit does not apply for those dusty dwarfs.

In  Figure \ref{Tseq_dusty},  we present  the full  dusty (AMES-Dusty)
limiting case from 2500 to  1500K.  Here the strong heating effects of
dust opacities prevent  the formation of methane bands,  and H$_2$O is
dissociated while producing a hotter water vapor opacity profile, much
weaker  and more  transparent  to radiation.   From  1700K, the  grain
opacity  profiles rapidly  dominate  the UV  to  red spectral  region,
smoothing out  the emergent flux into  a continuum. Only  the cores of
the strongest  atomic resonance  lines (Na~I~D and  K~I) can  be seen.
The  result  is  a   spectral  distribution  guetting  closer  to  the
equivalent blackbody  distribution of same  effective temperature (see
also Figure \ref{15CD}).  Note however, that Dusty models can never be
approximated  by blackbodies because  of the  important optical-to-red
dust  veiling, and  the strong  near-IR  water vapor  bands.  We  have
explored the effects of grain sizes on these models and found that for
grains  with sizes  in the  submicron to  micron range,  the increased
cross-sections are  compensated by the corresponding  reduction in the
number density of these grains given the conservation of the elemental
abundance.   For grains  with  sizes beyond  10  $\mu$m, an  increased
global opacity is found which produces even redder models.  But grains
are likely to be distributed in  a spectrum of sizes where the balance
between coagulation, sedimentation  and condensation decides the upper
limit of  the masses reached.  Preliminary  calculations (T.  Guillot,
private communication)  which will be published  separately show that,
when accounting for all the relevant processes, the grain sizes remain
in the submicron  range.  We are therefore confident  that the current
models  with grain  sizes  in  the submicron  range  do constitute  an
adequate full dusty limit for these dwarfs.

\subsection{Gravity}

In  Figure  \ref{Gseq_cond25} and  \ref{Gseq_cond05},  we explore  the
effects of surface gravity on AMES-Cond models with $\teff = 2500$ and
500K.   At 2500K,  gravity  sensitivity is  essentially noticeable  in
hydride bands  (CaH at 0.624 and  0.639 and FeH at  0.98~$\mu$m) while
the  oxide  bands  (TiO  and  VO)  form too  high  in  the  atmosphere
($\tau_{\rm  std}=10^{-4}$)   to  be  affected   except  in  interband
pseudo-continuum  regions.   Effects   of  gravity  are  however  more
extensive in the  strength of atomic lines (essentially  K~I and Ti~I)
at the peak of the  spectral distribution (1.05 to 1.3~$\mu$m), and in
the red wing  of the water vapor bands  as well as in the  CO bands at
2.3 to 2.4~$\mu$m.  In the 500K case, H$_2$O bands are only moderately
affected by  the gravity change, while the  optical continuum opacity,
provided by  the van der Waals  wings of the Na~I~D  and K~I resonance
doublets, is reduced by  nearly a factor of 10 in the  500K in the low
gravity  case  in  response  to   the  drop  in  pressure.   The  most
interesting feature  is the enhanced sensitivity of  the $K$-band flux
at 2.2~$\mu$m  to gravity.  As  already pointed out  by \cite{gl229b},
this feature can be used to disentangle temperature, age and mass of a
brown dwarf  or planet independently.   This trend is observed  in the
entire regime  from 1500K to 300K,  and provides a useful  tool in the
analysis of  free floating methane dwarfs such  as discovered recently
by \cite{Strauss99}  and \cite{Tsvetanov00}.  The CH$_4$  and CO bands
are also sensitive to gravity in this regime.

One  more gravity  indicator should  be the  slope and  height  of the
$Z$-band  flux  between the  core  of  the  K~I resonance  doublet  at
$\lambda$7687,7701\AA\  to 1.1~$\mu$m  compared to  the height  of the
$J$-band flux peak.   However, for the reasons mentioned  above, it is
difficult to quantify this effect on the basis of the present models.

Surface gravity  effects have also  been explored for the  fully dusty
case  (AMES-Dusty models)  at $\teff  = 2000$  and 1500K  (see Figures
\ref{Gseq_dusty20}  and \ref{Gseq_dusty15}).  In  the 2000K  case, the
pseudo-continuum  formed  by  saturated  bands  of  TiO  bluewards  of
0.75~$\mu$m  is fainter  and flatter  at  reduce gravity.   This is  a
result of the  cooler temperatures prevailing in the  outskirts of the
photosphere  at  reduced  gravity.   This explains  and  supports  the
conclusions of \cite{1996ApJ...469..706M}  who noticed a similar trend
comparing young red dwarfs of the Pleiades cluster to presumably older
field M dwarfs.  Previous M dwarf model atmospheres \cite[]{MDpap} did
not show such  a sensitivity due to the  overestimated blocking caused
by  straight  mean  and  JOLA  opacities.  To  longer  wavelengths  an
important veiling  provided by the  dust covers the 0.7  to 1.3~$\mu$m
region in  the high gravity  model.  Atomic lines and  molecular bands
are generally quite  sensitive to gravity change in  this range, while
the near-infrared  water vapor bands are also  affected markedly (much
more than  in the grainless 2500K case  discussed above).  Especially,
collision  induced H$_2$  absorption cutting  the flux  in  a negative
slope at  2.2-2.3~$\mu$m, as  well as the  CO bands redwards  of this,
makes  again the shape  of the  K band  spectrum an  excellent gravity
indicator.   A   similar  behavior   is  observed  at   1500K  (Figure
\ref{Gseq_dusty15}), where however the dust is now so strong that most
features, except H$_2$O band troughs, are no longer seen.

\subsection{Comparing limits}

Figures  \ref{20CD}  and  \ref{15CD}  compare  the  full-settling  and
full-dusty limits for  $\teff = 2000$ and 1500K  respectively.  In the
2000K case,  the presence of dust  opacities simply has  the effect to
veil the  optical to  red spectral region,  while the  additional heat
causes flux redistribution to  the infrared.  This $\teff$ and gravity
is typical of dusty red and  brown dwarfs.  In the 1500K case however,
the dust  opacity profiles are  blocking nearly all flux  bluewards of
1.0~$\mu$m, and  only the water vapor bands  are still distinguishable
in the AMES-Dusty model.  The AMES-Cond models, on the other hand, are
very transparent  since all trace  of TiO, VO,  CaH, MgH and  FeH have
vanished through condensation to dust grains.  And since more flux can
escape from short wavelengths, the  upper atmosphere is cooler and the
water bands stronger.

\begin{figure}[]
\psfig{file=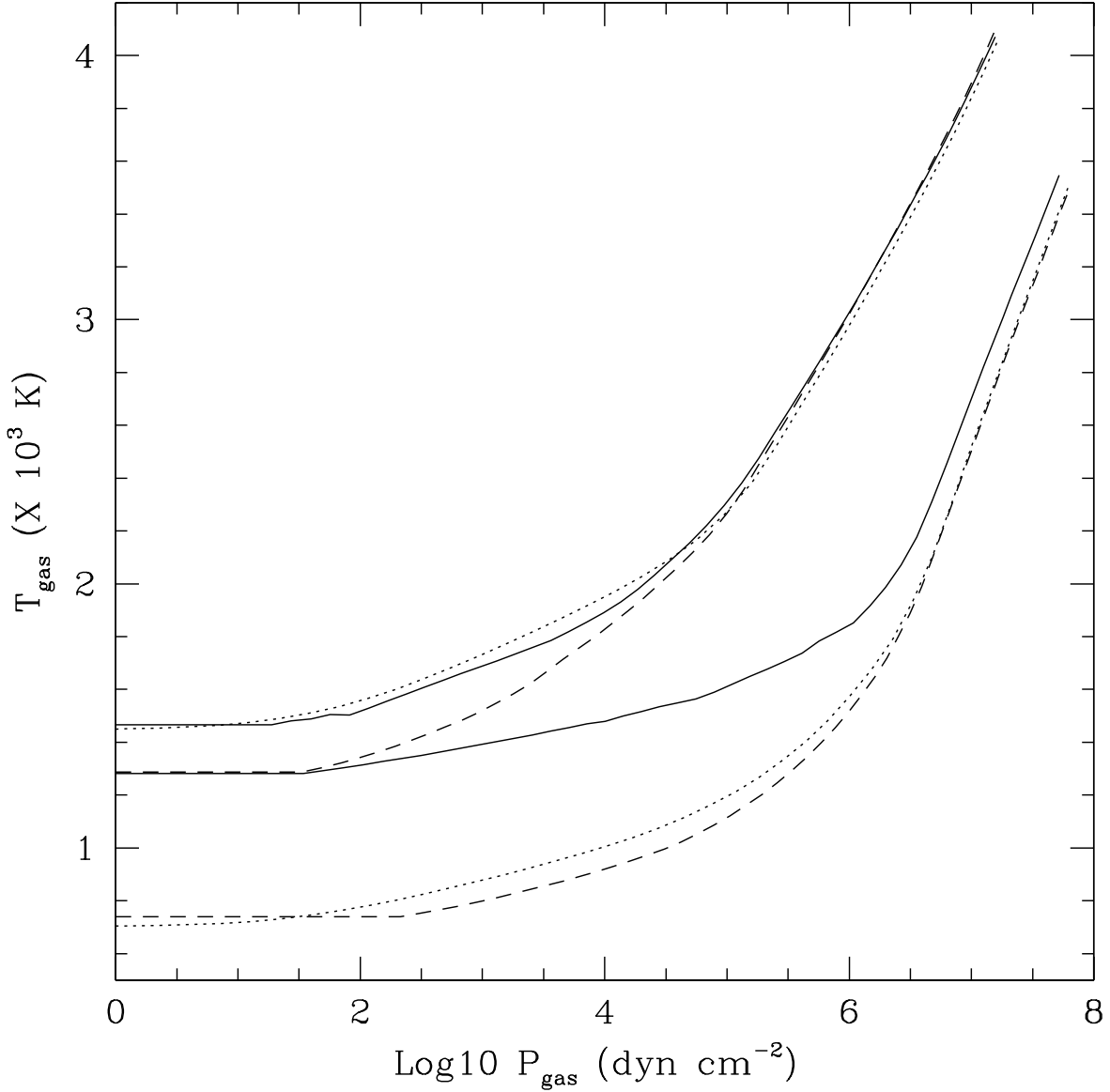,width=\hsize,angle=0}
\caption[]{\label{heat} Thermal structures of models with T$_{\rm eff}
= 2800$K and 1800K, logg=5.0, and solar metallicity for three types of
models:  (1) the  standard NextGen  models treated  in gas  phase only
(dotted line); (2) the  AMES-Dusty models assuming a full distribution
of the dust (full line),  and; (3) the AMES-Cond models including dust
in the CE but ignoring  their opacities (dashed line).  All models are
converged.  Note  that the  NextGen models use  a different  source of
water vapor opacity (see text).}
\end{figure}

\begin{figure}[]
\psfig{file=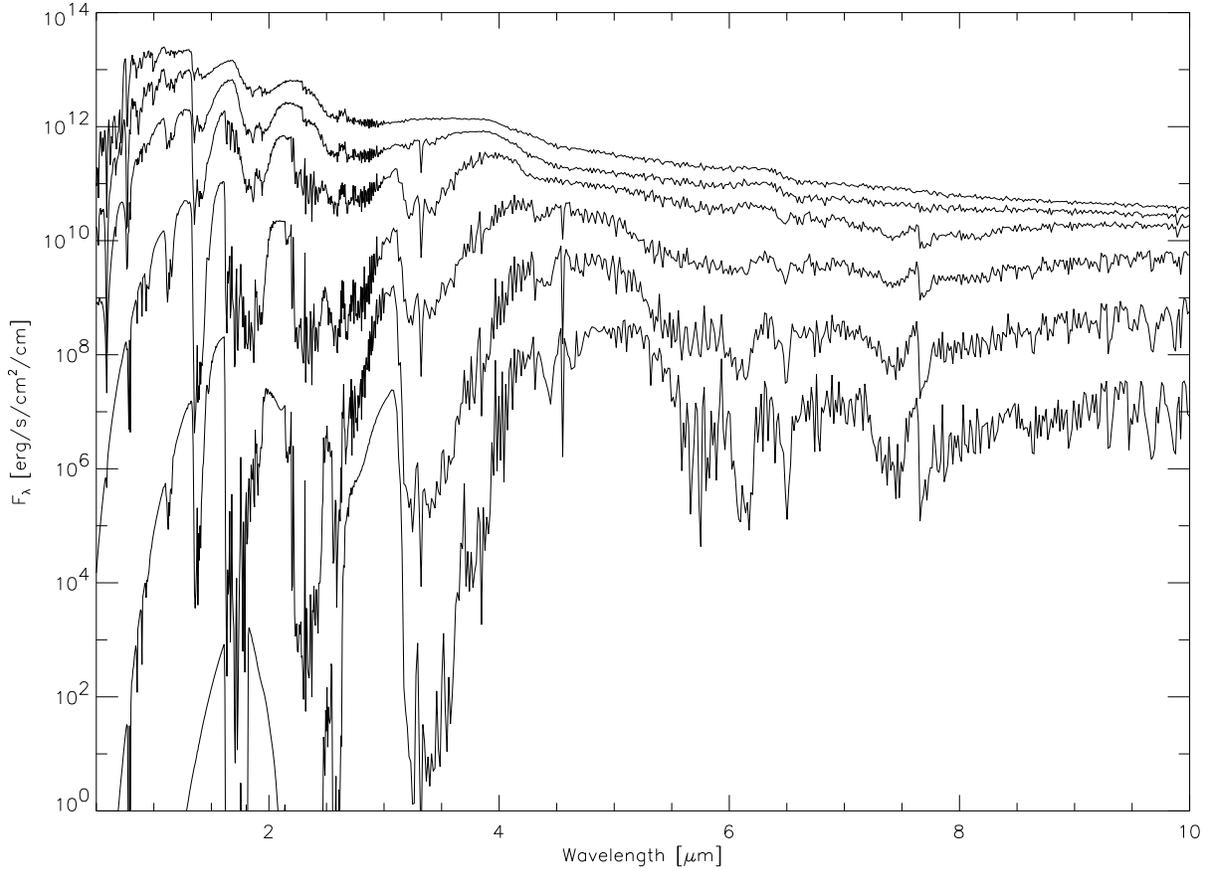,width=\hsize,angle=90}
\caption[]{\label{Tseq_cond}Spectral sequence  of brown dwarfs  to EGP
model  atmospheres in  the total  settling  (AMES-Cond) approximation.
From top  to bottom: T$_{\rm eff}=  2500, 1900, 1300,  700, 400$K, and
200K.  The gravity is fixed to $\log g=5.0$.  These models (AMES-Cond)
assume  complete  settling of  the  grains  (i.e.   neglects all  dust
opacity).   The spectral  resolution has  been reduced  from  2\AA\ to
30\AA\ by boxcar smoothing in  order to make comparison of the spectra
easier.   We observe  that CH$_4$  bands already  develop at  2000K in
these  extremely  transparent and  cool  AMES-Cond atmospheres.   They
gradually  replace water  vapor bands  while H$_2$O  condenses  to ice
below 300K.}
\end{figure}

\begin{figure}[]
\psfig{file=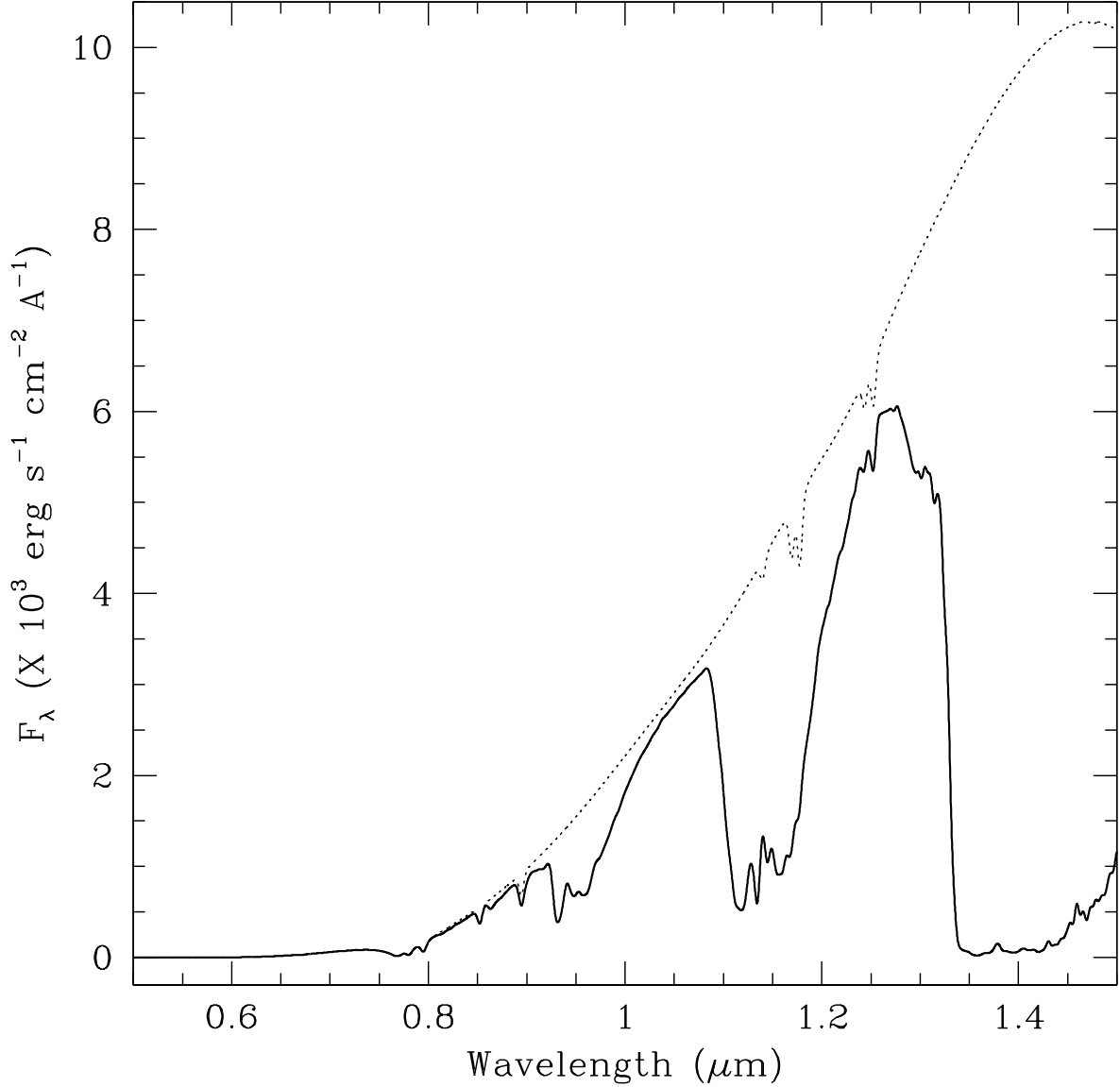,width=\hsize,angle=0}
\caption[]{\label{vdw}To isolate atomic features we compare a $\teff =
1000$K  AMES-Cond  model  (full  line)  with a  spectrum  obtained  by
neglecting all  molecular lines (dotted  lines).  The pseudo-continuum
is essentially formed by the van der Waals wings of the Na~I~D and K~I
resonance       doublets      at       $\lambda$5891,5897\AA\      and
$\lambda$7687,7701{\AA}.   Weaker  lines  of Rb~I  ($\lambda$7802  and
7949{\AA}),    Cs~I   ($\lambda$8523    and   8946{\AA}),    and   K~I
($\lambda$11693,11776  and $\lambda$12436,12525{\AA})  are  also seen.
The background flux (obtained  by neglecting both atomic and molecular
lines, not shown)  lies outside the plot!  The  spectral resolution is
reduced to 30\AA\ for this illustration.}
\end{figure}

\begin{figure}[]
\psfig{file=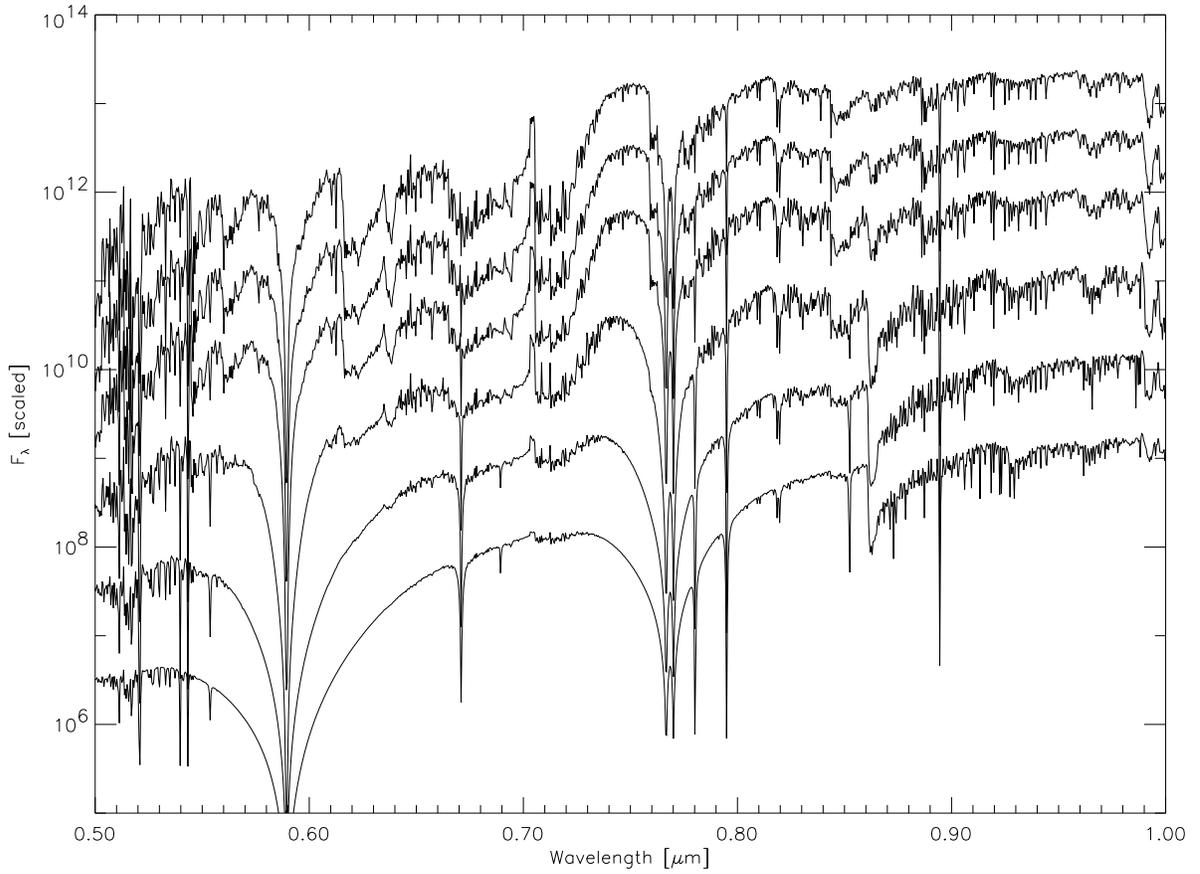,width=\hsize,angle=90}
\caption[]{\label{Tseq_cond_optH}Same as Figure \ref{Tseq_cond} in the
optical  to  near-red spectral  range.   Here  the  spectra have  been
arbitrarily scaled to facilitate  the comparison.  From top to bottom:
T$_{\rm eff}= 2500, 2400, 2300,  2000, 1700$, and 1500K.  The spectral
resolution is 2\AA.}
\end{figure}

\begin{figure}[]
\psfig{file=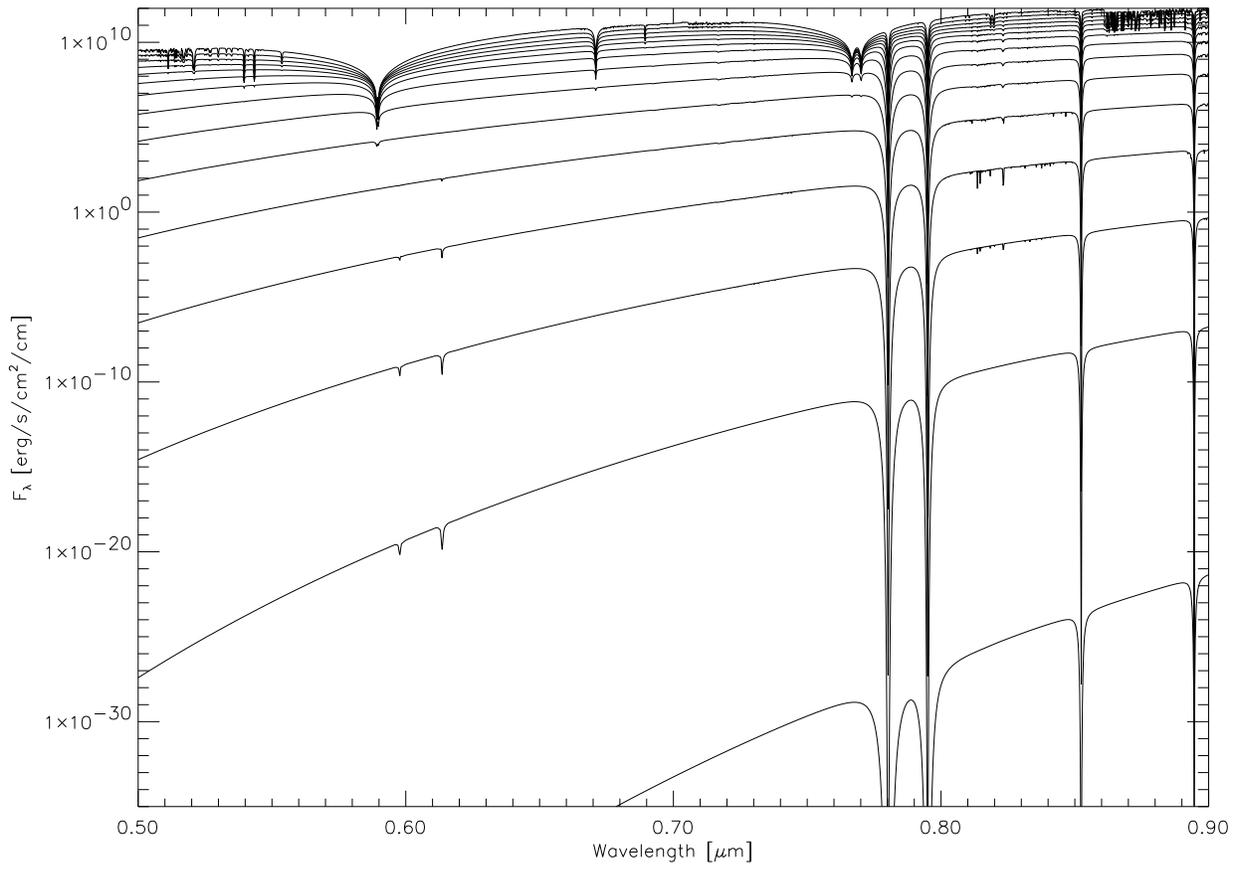,width=\hsize,angle=90}
\caption[]{\label{Tseq_cond_optC}Same  as  Figure \ref{Tseq_cond_optH}
for models from 1500K to 100K at 100K steps.}
\end{figure}

\begin{figure}[]
\psfig{file=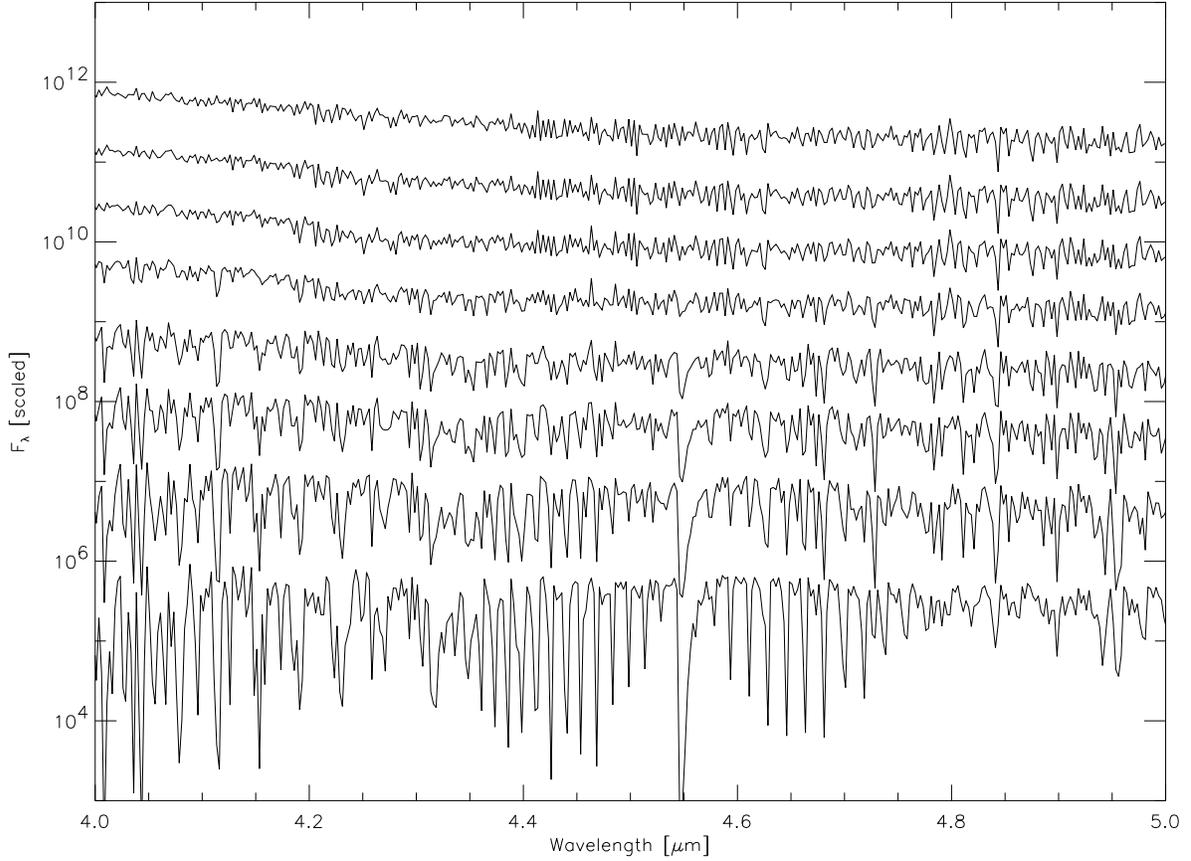,width=\hsize,angle=90}
\caption[]{\label{Tseq_cond_co}Same   as  Figure  \ref{Tseq_cond_optH}
where we zoom in on the  CH$_3$D band system at 4.55~$\mu$m.  From top
to bottom: T$_{\rm eff}= 2000,  1700, 1500, 1300, 1000, 800, 600$, and
400K.  The CH$_3$D band appears at  1000K at this gravity and for this
dust treatment limit.  Note that  $\teff = 2000$K dwarfs are dusty and
that, though  they don't appear  here, CO bands at  4.67~$\mu$m easily
{\bf are} detectable in these hotter atmospheres.}
\end{figure}

\begin{figure}[]
\psfig{file=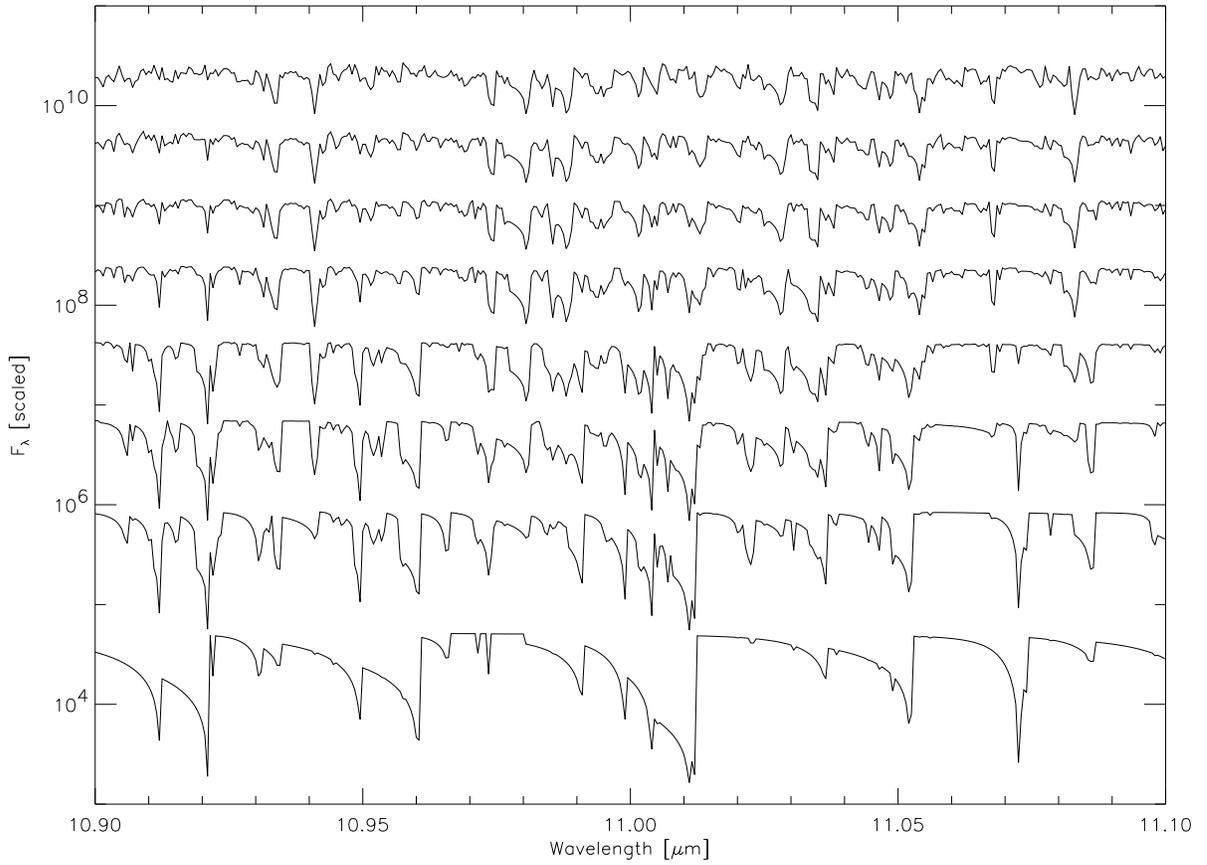,width=\hsize,angle=90}
\caption[]{\label{Tseq_cond_nh3}Same  as  Figure  \ref{Tseq_cond_optH}
where  we zoom in  on the  NH$_3$ band  system at  11.012~$\mu$m.  The
ammonia  band  system  appears  at  1000K, along  with  several  other
molecular lines  essentially due to methane.   The spectral resolution
is 5\AA.}
\end{figure}

\begin{figure}[]
\psfig{file=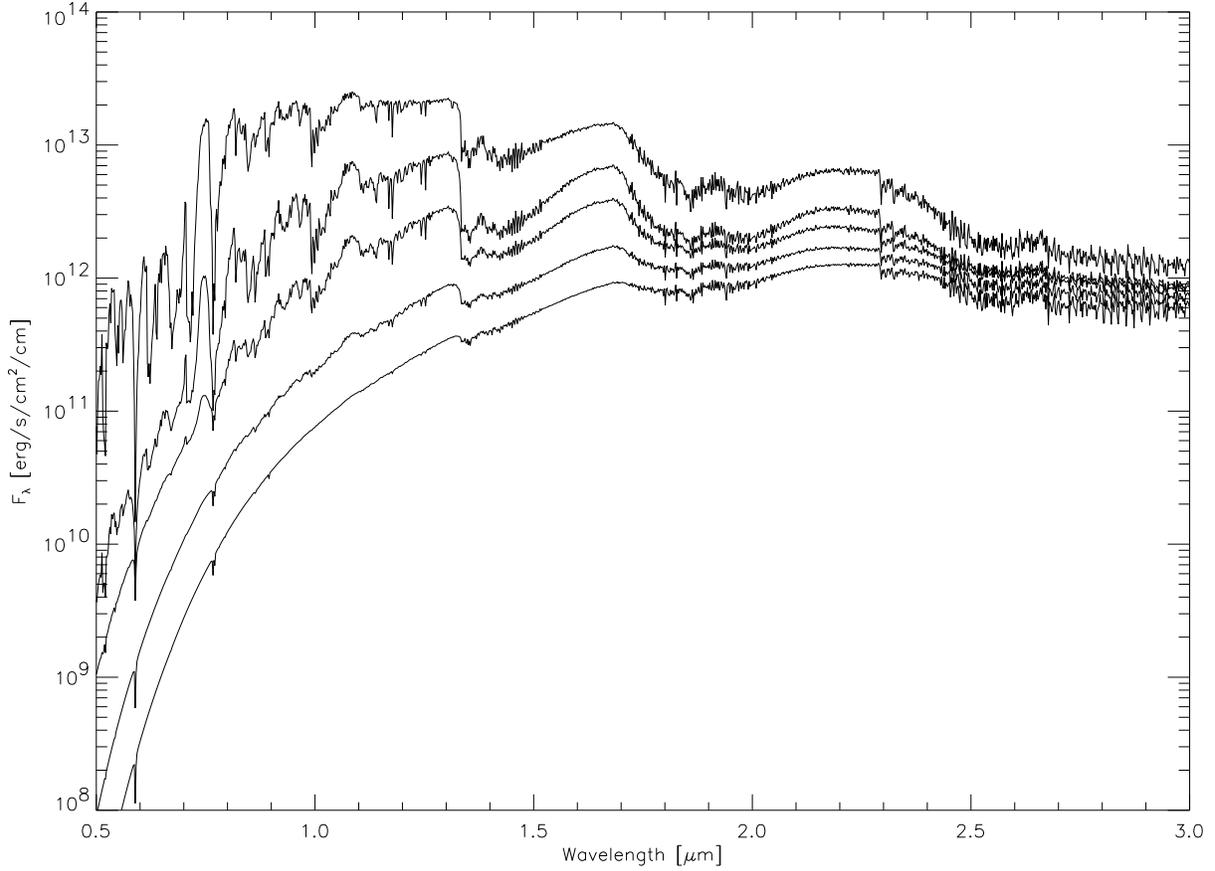,width=\hsize,angle=90}
\caption[]{\label{Tseq_dusty}Same  as  Figure  \ref{Tseq_cond} in  the
full dusty  (AMES-Dusty) limiting case.   From top to  bottom: T$_{\rm
eff}= 2500,  2000, 1800, 1600$K, and  1500K.  The gravity  is fixed at
$\log  g= 5.0$.   Here the  strong heating  effects of  dust opacities
prevent the  formation of methane  bands, and dissociate  H$_2$O while
producing  a  hotter water  vapor  opacity  profile,  much weaker  and
transparent  to radiation.   From  1700K, the  grain opacity  profiles
rapidly  dominate the  UV to  red spectral  region, smoothing  out the
emergent flux  into a continuum where  only the core  of the strongest
atomic resonance lines (Na~I~D and K~I doublets) are seen. }
\end{figure}

\begin{figure}[]
\psfig{file=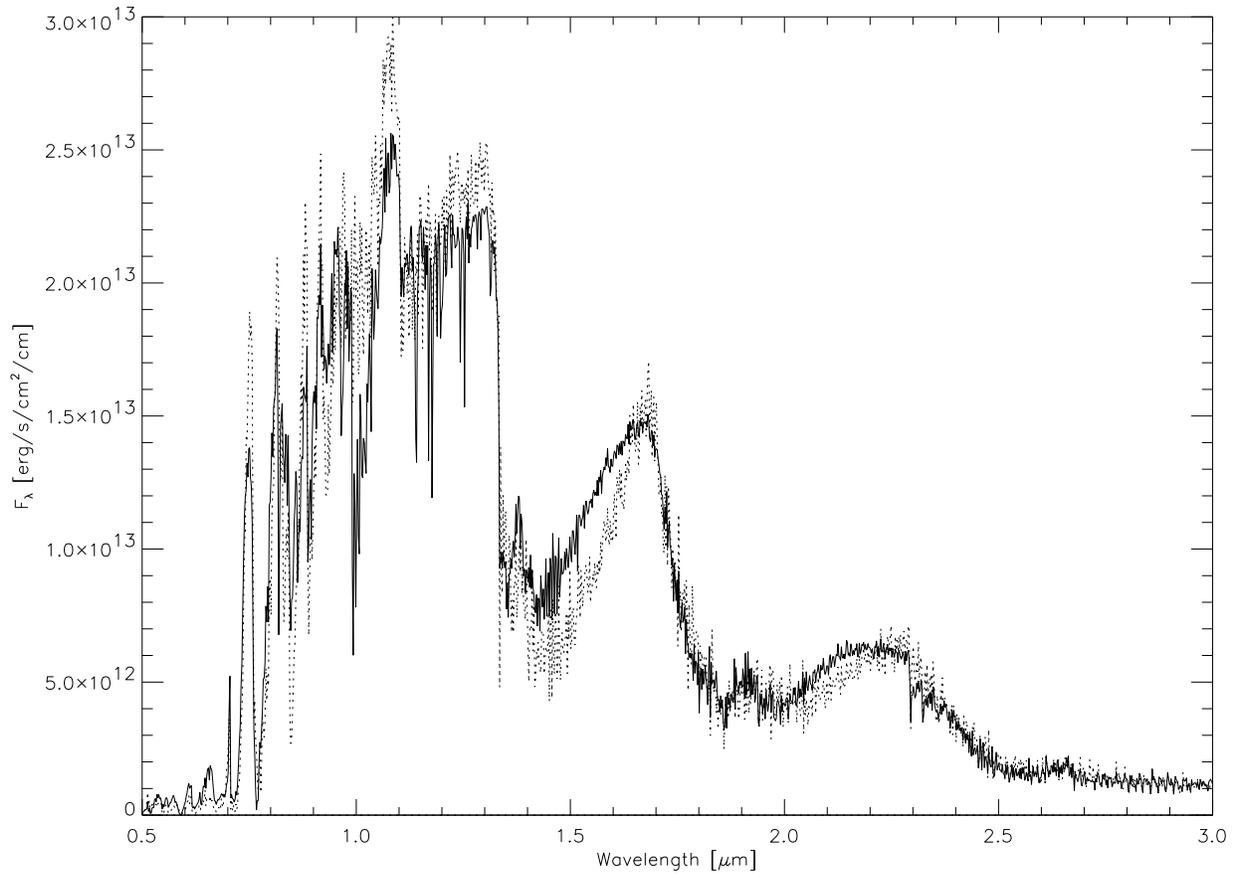,width=\hsize,angle=90}
\caption[]{\label{Gseq_cond25}Two $\teff =  2500$K AMES-Cond models of
different surface gravity are compared: (1) $\log g= 5.5$ (full line),
(2) $\log  g= 2.5$ (dotted  line).  The  spectral resolution  has been
reduced  from 2\AA\ to  10\AA\ by  boxcar smoothing  in order  to make
comparison of the spectra easier.}
\end{figure}

\begin{figure}[]
\psfig{file=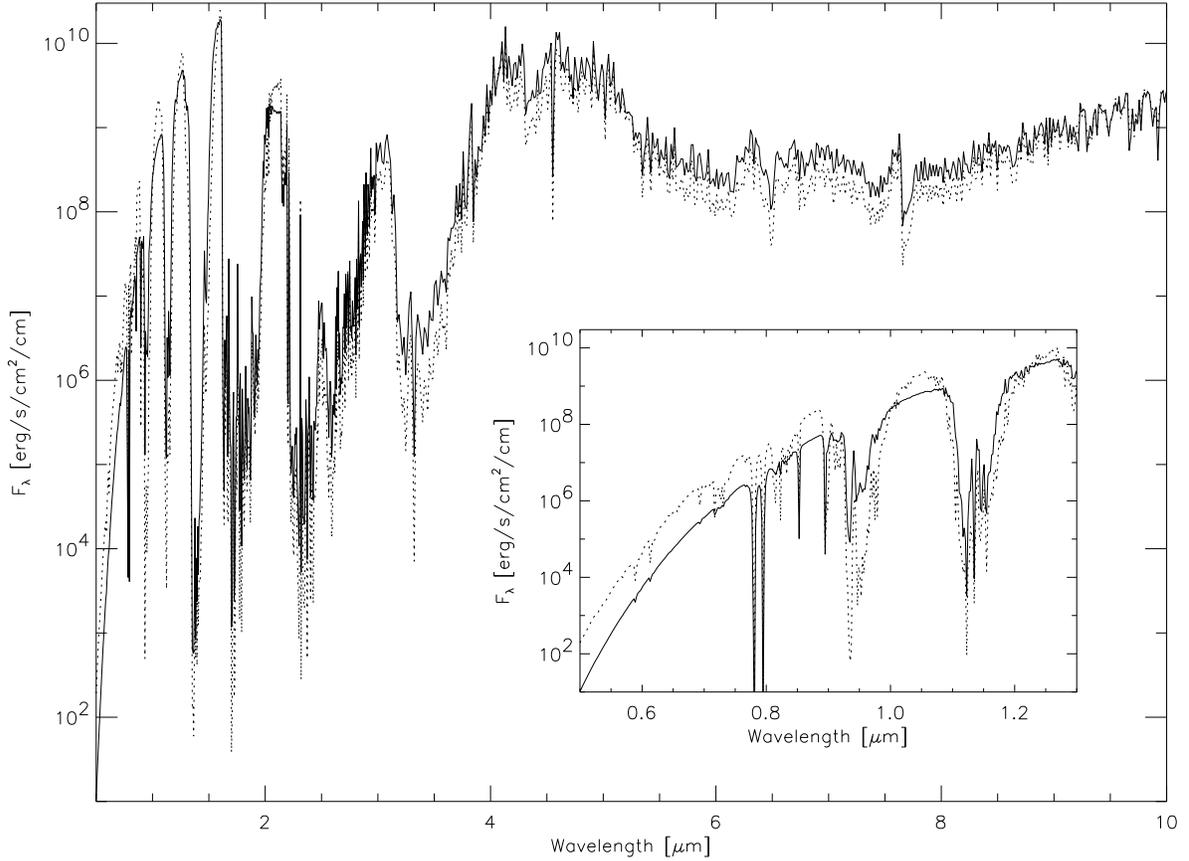,width=\hsize,angle=90}
\caption[]{\label{Gseq_cond05}Same  as  Figure  \ref{Gseq_cond25}  for
$\teff =  500$K AMES-Cond models, and  (1) $\log g=  4.0$ (full line),
(2) $\log  g= 2.5$ (dotted  line).  Here  the spectral  resolution has
been reduced  to 30\AA\ by boxcar  smoothing.  In the inset  we show a
zoom of the optical to  red spectral regime, were we distinguish water
vapor bands at 0.93, 0.95  and 1.12 $\mu$m, Cs~I resonance transitions
at  0.86 and  0.89 $\mu$m,  the K~I  resonance doublet  at  0.77, 0.79
$\mu$m, and the cores of a  few other lines such as the Na~I~D doublet
bluewards  of  0.75  $\mu$m.   While molecular  bands  are  moderately
affected by the gravity change,  the optical background opacity due to
the wings of the Na~I~D and K~I doublets is reduced by nearly a factor
of 10 in  the low gravity model.  The latter model  is more typical of
low mass brown dwarfs and jovian planets.}
\end{figure}

\begin{figure}[]
\psfig{file=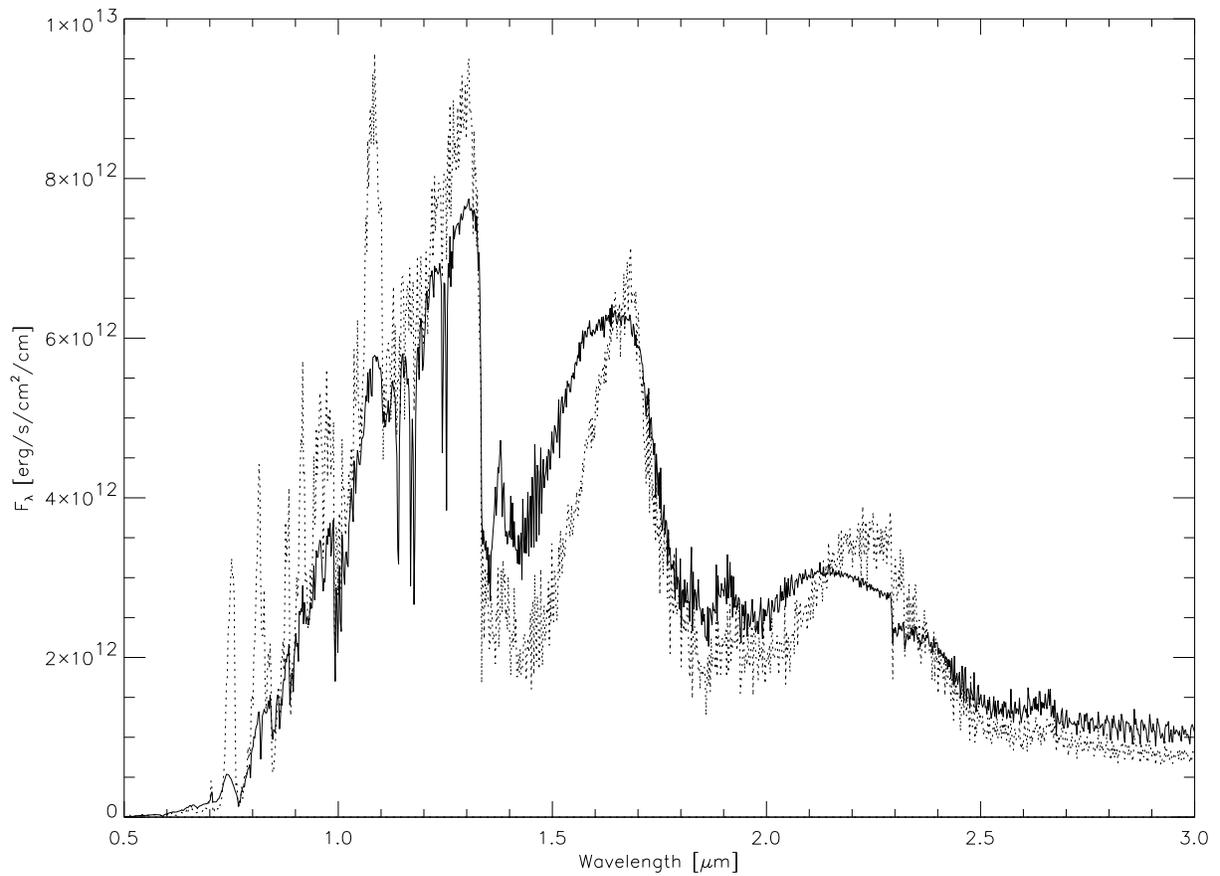,width=\hsize,angle=90}
\caption[]{\label{Gseq_dusty20}Two  $\teff = 2000$K  AMES-Dusty models
of different  surface gravity  are compared: (1)  $\log g=  6.0$ (full
line), (2) $\log  g= 3.5$ (dotted line).  The  spectral resolution has
been reduced to 10\AA\ by boxcar smoothing.}
\end{figure}

\begin{figure}[]
\psfig{file=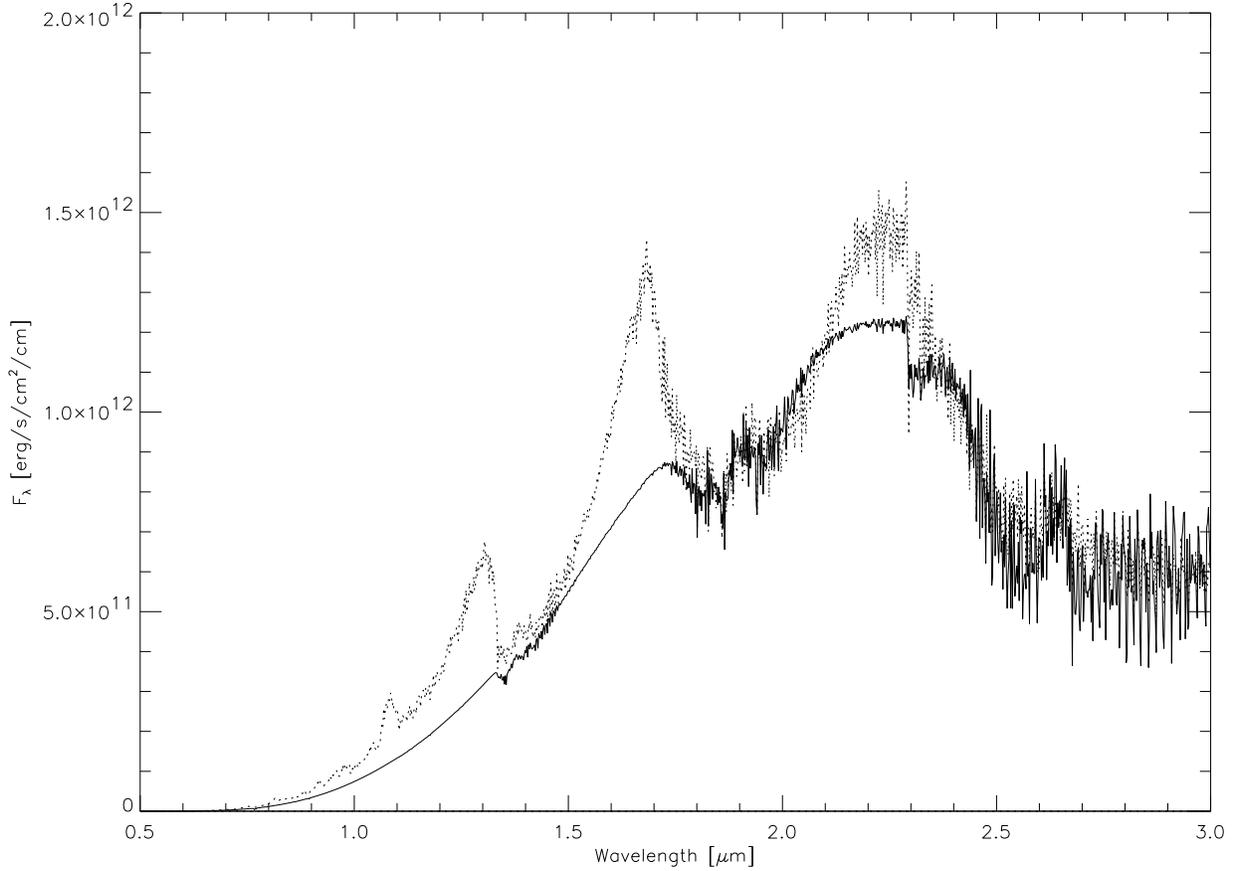,width=\hsize,angle=90}
\caption[]{\label{Gseq_dusty15}Same  as Figure  \ref{Gseq_dusty20} for
$\teff =  1500$K AMES-Dusty models.   While gravity effects  are quasi
nonexistent  redwards of  2.5  $\mu$m,  they are  quite  large in  the
optical  to  red spectral  region,  mainly  as  a result  of  enhanced
efficiency  of  dust  grain  formation  at the  higher  pressures  and
densities of  high gravity  atmospheres.  A $\log  g$-value of  5.5 is
typical of most field brown dwarfs discovered since 1996.}
\end{figure}

\begin{figure}[]
\psfig{file=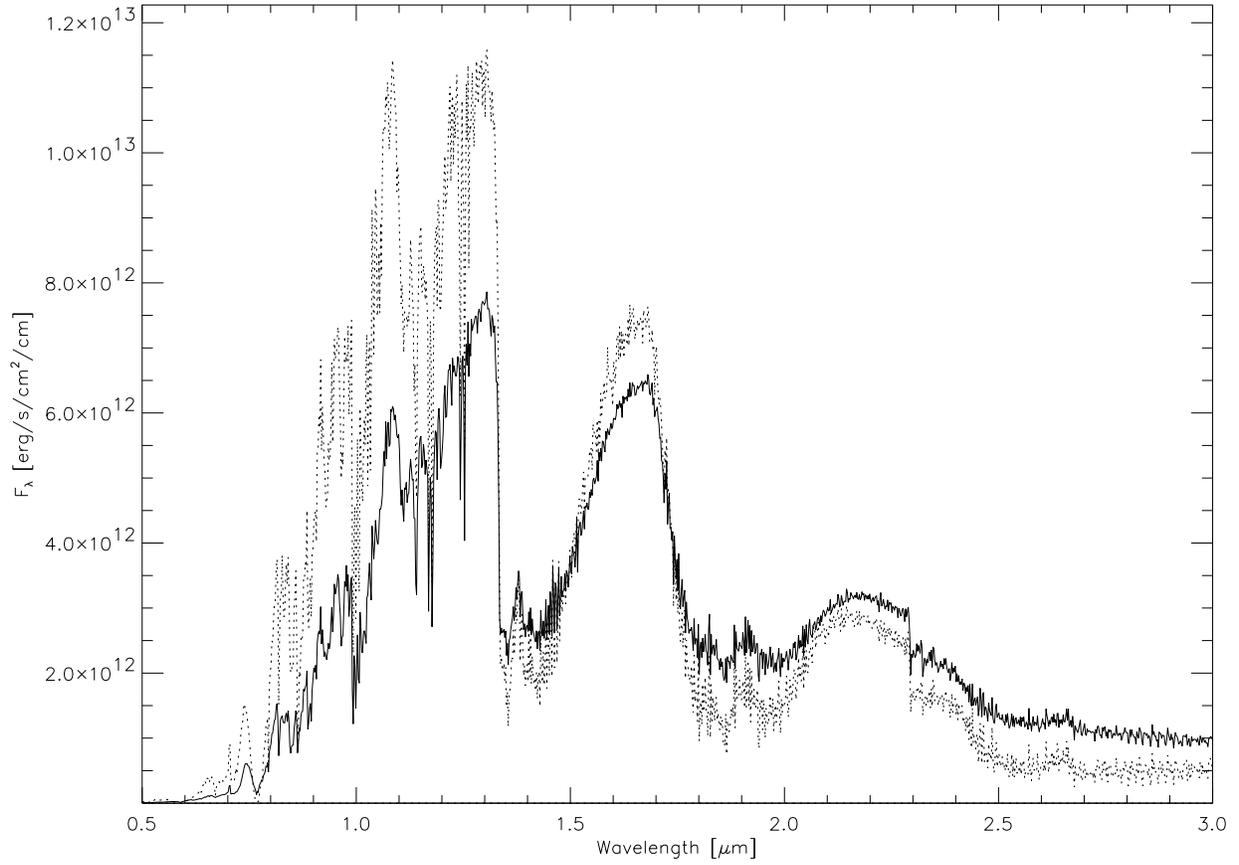,width=\hsize,angle=90}
\caption[]{\label{20CD}Two $\teff  = 2000$K models with  $\log g= 5.5$
are  compared to illustrate  the difference  between our  two limiting
cases:  (1) AMES-Dusty  with full  dust opacity  (full line),  and (2)
AMES-Cond with  full gravitational  settling (no dust  opacity, dotted
line).  }
\end{figure}

\begin{figure}[]
\psfig{file=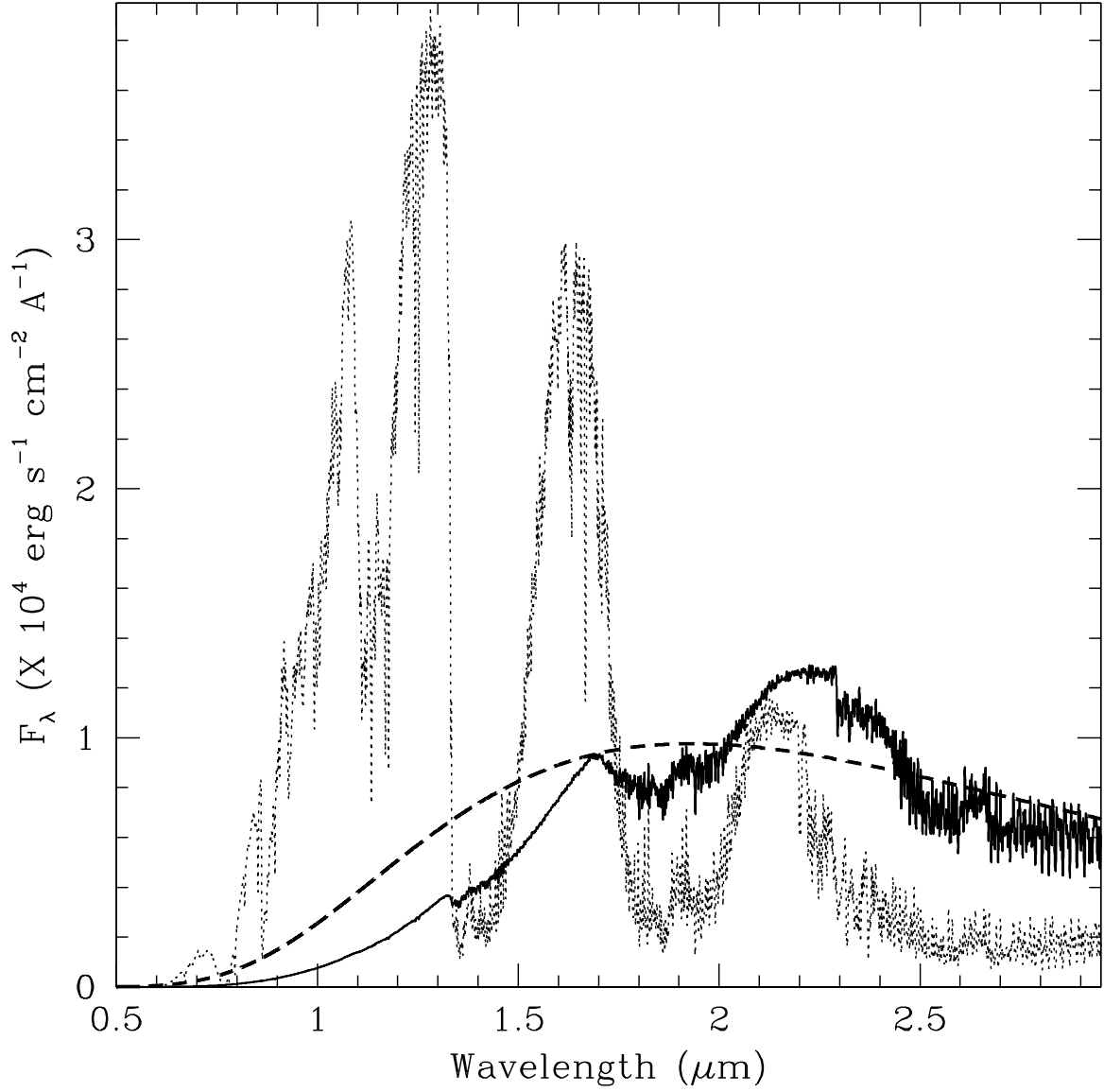,width=\hsize,angle=0}
\caption[]{\label{15CD}Same as  Figure \ref{20CD} for  $\teff = 1500$K
models  with  $\log g=  5.0$.   A  1500K  blackbody (dashed  line)  is
overplotted for comparison.  }
\end{figure}

As  mentioned earlier,  one  of  the most  important  effects of  dust
extinction  in  the  photospheres  of  red and  brown  dwarfs  is  the
resulting  heating  of  the   outer  atmospheric  layers.   In  Figure
\ref{heat},  we compare  the thermal  structures of  our  two limiting
cases to NextGen  models for two $\teff$: one  typical of M~dwarfs and
young/massive brown dwarfs ($2800$K), the other typical of the reddest
L~dwarfs ($1800$K).  In  the former case, the cloud  layers form above
the photosphere, as can be  seen from Figure \ref{TiOeos26}, such that
heating takes place above the line forming region, leaving the thermal
structure little affected  by the dust.  At 1800K,  on the other hand,
the clouds form deep into the photosphere (see Figure \ref{TiOeos18}).
The line  forming region is  therefore heated up  by as much  as 500K,
while the internal layers only warm  up by less than 80K.  This effect
is  enough to  dissociate  water vapor  by  nearly 50\%  in the  outer
layers,  leaving   a  far  shallower   structure  over  most   of  the
photosphere.  It is  interesting to note that Cond  models cooler than
about  $\teff   =  1800$K,   have  similar  thermal   structures  than
corresponding NextGen models.

\begin{figure}[]
\psfig{file=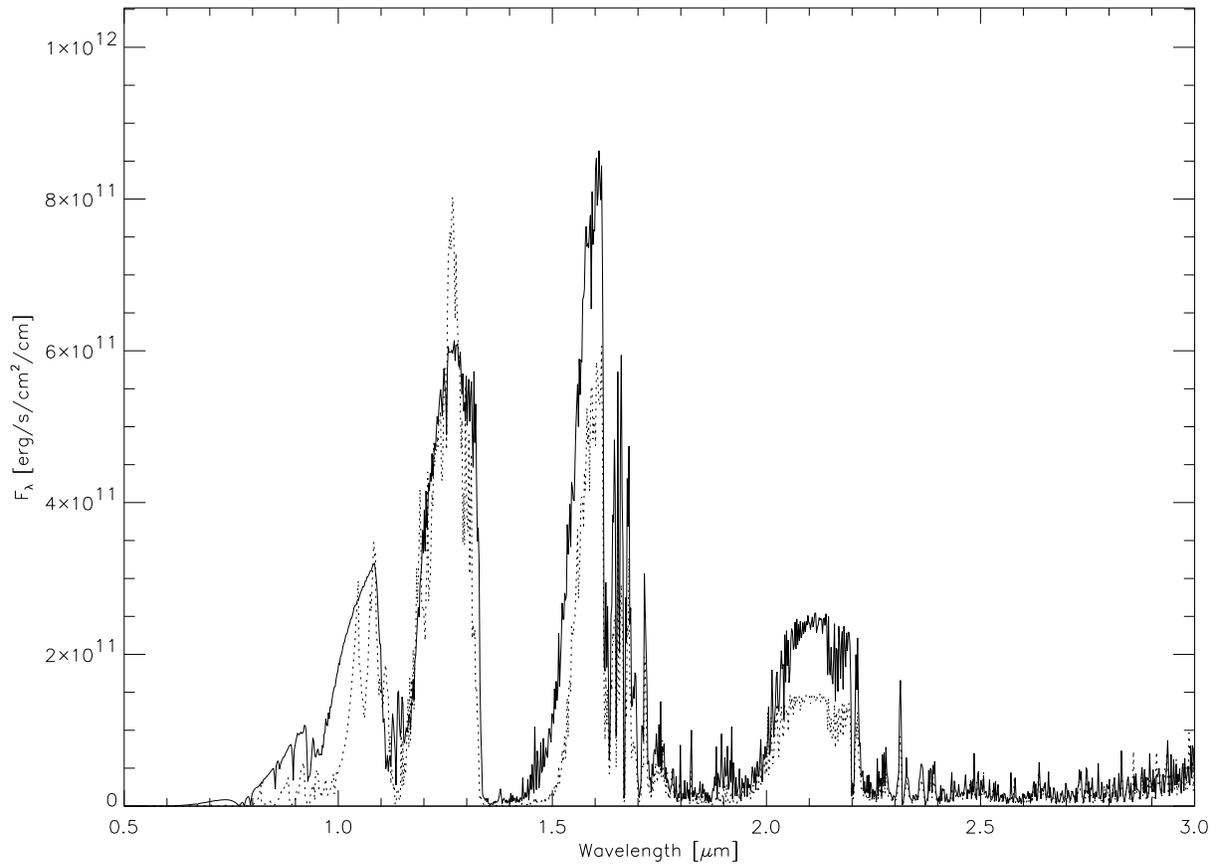,width=\hsize,angle=90}
\caption[]{\label{CNG10}The  AMES-Cond model for  $\teff =  1000$K and
$\log  g=   5.0$  (full  line)   is  compared  to   the  corresponding
\cite{gl229b} model (dotted line) used in their analysis of the Gl229B
spectrum.  }
\end{figure}

This  similarity of  the  thermal structures  translates into  similar
synthetic  spectra.  We compare  the present  AMES-Cond models  to our
earlier brown  dwarfs models (NextGen)  as of \cite{gl229b}  in Figure
\ref{CNG10} for the $\teff = 1000$K case.  Although the NextGen models
used   in  our   1996  publication   were  constructed   without  dust
condensation, the molecular  bands of TiO, VO and  FeH were completely
crushed  by  the wings  of  the alkali  lines  and  the results  quasi
independent  of  condensation.  The  models  remain  sensitive to  the
physics  mainly  in the  inter-band  regions  at  1.0, 1.25,  1.6  and
2.2~$\mu$m  which probe,  by  their relative  brightness, the  thermal
structure of the atmosphere at various depths.

\section{\label{color}Colors}

The models  are most  readily compared to  large samples of  stars and
brown  dwarfs in  color-magnitude  diagrams.  The  standard system  of
broadband  colors  is sufficiently  constraining  when evaluating  the
accuracy of  the models.  This  is because most spectral  features are
several  thousands  of angstroems  wide,  and  the remaining  emission
windows are well  sampled by each bandpass: $Z$  at $1.0~\mu$m, $J$ at
$1.3~\mu$m, $H$  at $1.6~\mu$m, $K$ at $2.2~\mu$m,  $M$ at $4.5~\mu$m,
and $N$  at $10~\mu$m.   Methane bands appear  in brown  dwarfs cooler
than about 1700K at 1.7, 2.4 and 3.3~$\mu$m, reducing the flux sampled
by   the   $H$,   $K$   and   $L$'   bandpasses   respectively.    The
pressure-induced H$_2$ opacity, on  the other hand, depresses the flux
in the  $K$ bandpass in  the coolest brown dwarfs  and low-metallicity
dwarf stars.

\begin{figure}[]
\psfig{file=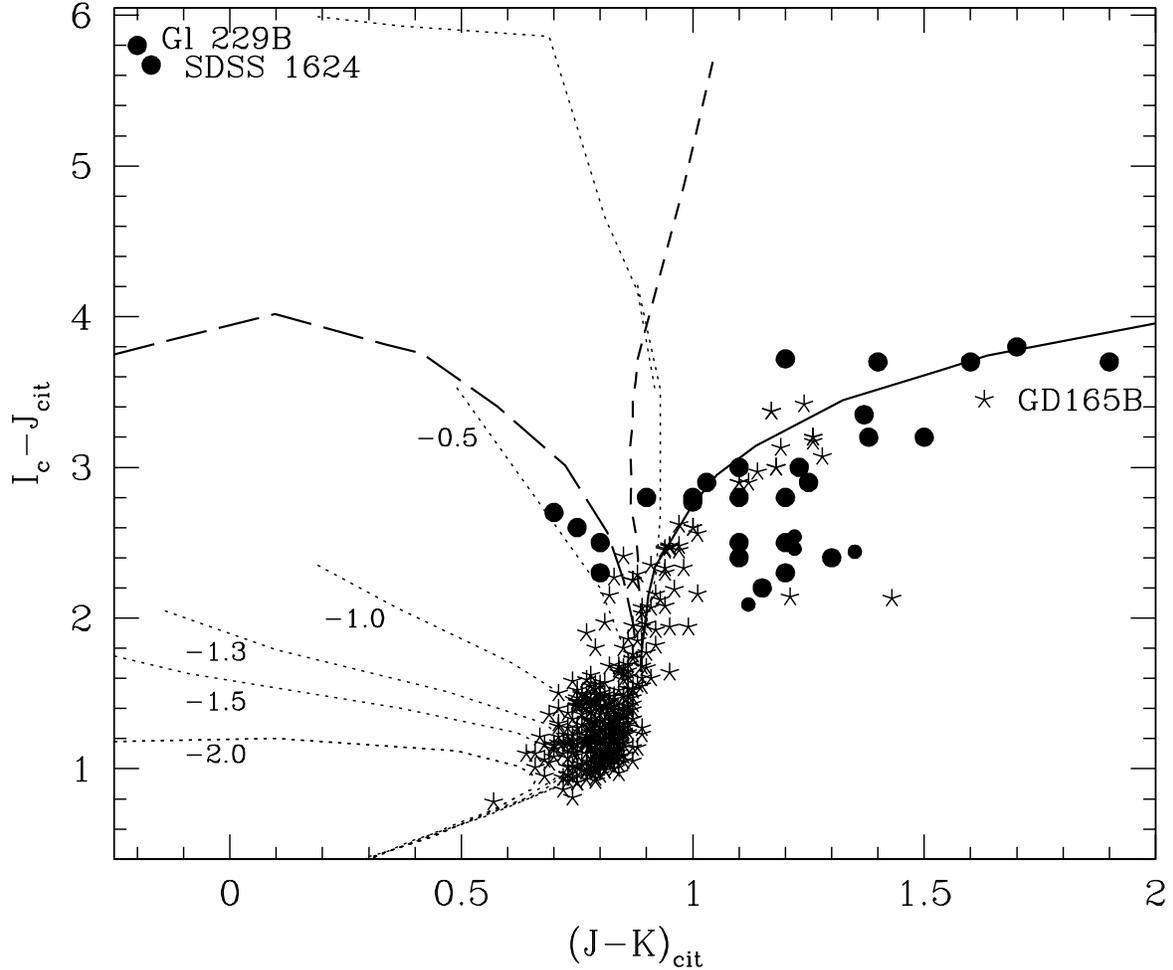,width=\hsize,angle=0}
\caption[]{\label{IJK}The 10 Gyr  NextGen isochrones (dotted), and the
$\log g=5.0$ locii of the  Cond (short dashed) and Dusty (full) models
are compared to the photometric  observations of field stars and brown
dwarfs,  and to  Pleiades objects  including the  brown  dwarfs PPl15,
Teide1  and Calar3  (star and  filled  circle symbols).   The field  T
dwarfs Gliese  229B and SDSS1624 are also  shown.  Unresolved binarity
is reflected in this diagram by  a red excess in $J-K$.  Note that the
Cond and Dusty models have been  shifted in $J-K$ by +0.15 in order to
eliminate water  opacity source effects in this  comparison.  The Cond
models are computed (i) with the line wings coverage value of 5000\AA\
(long-dashed  line), and  (ii) with  a maximum  coverage  of 15000\AA\
(short-dashed line) to illustrate the  impact of this parameter on the
Cond models.}
\end{figure}

We  have computed synthetic  $UBVRIJHKLL'M$ magnitudes  by integrating
our  model spectra  according to  the photon  count prescription  at a
wavelength step  of 1{\AA}.  We  have adopted the filter  responses by
\cite{Bes88}  and \cite{Bes90}, bringing  our synthetic  photometry on
the Cousins  and Johnson-Glass system.  Transformations to  the CIT or
to other  systems are readily  obtained from \cite{leggett92}.   As in
previous papers, we  used the energy distribution of  Vega as observed
by  \cite{Hayes75,Hayes85} and \cite{Mount85}  to provide  an absolute
calibration.  Zero  magnitudes and colors  are assumed for  Vega.  The
grainless                NextGen               models               of
\cite{gl229b,araa,ng-hot,clusterpap,BCAH98}, as  well as the AMES-Cond
and  AMES-Dusty models  from this  work are  compared to  the observed
stellar    and    brown    dwarfs   samples    of    \cite{leggett92},
\cite{1993ApJ...414..279T},                 \cite{1994AJ....107..333K},
\cite{1995ApJ...441L..47I},       \cite{1997A&A...327L..25D}       and
\cite{1997A&A...323..105Z} in  Figure~\ref{IJK}.  Please note  that we
have applied here a +0.18 dex  shift in $J-K$ to the current models to
match the  position of the NextGen  models in the  non-dusty regime in
order to isolate the dust  effects.  This offset of the current models
to the blue of the earlier  NextGen models is due to some inaccuracies
of  the   NASA-Ames  H$_2$O  opacity  database   in  describing  these
relatively hot  atmospheres (see Allard, Hauschildt  and Schwenke 2000
for details\nocite{AMESpap}).

These  colors   are  interesting  as  they   have  helped  distinguish
interesting brown  dwarf candidates from the databases  of large scale
surveys such as  DENIS and 2MASS, and in  obtaining an appreciation of
the  spectral sensitivity  needed  to detect  new  brown dwarfs.   The
methane  bands  cause  the  $J-K$   colors  of  brown  dwarfs  to  get
progressively bluer with  decreasing mass and as they  cool over time.
Yet their $I-J$ colors remain  very red which allows us to distinguish
them  from hotter  low-mass  stars, red  shifted  galaxies, red  giant
stars, and even  from low metallicity brown dwarfs  that are also blue
due  to  pressure-induced  H$_2$  opacities  in  the  $K$  bandpasses.
Fortunately, grain formation  and uncertainties in molecular opacities
are  far  reduced  under  low metallicity  conditions  ([M/H]$<-0.5$).
Therefore,  model atmospheres  of metal-poor  subdwarf stars  and halo
brown dwarfs are  free of uncertainties on the  dust compared to their
metal-rich  counterparts.   This   has  been  nicely  demonstrated  by
\cite{clusterpap}  who  reproduced   closely  the  main  sequences  of
globular  clusters  ranging in  metallicities  from  [M/H]$= -2.0$  to
$-1.0$, as well  as the sequence of the  \cite{monet92} halo subdwarfs
in color-magnitude diagrams.

As can be seen  from Figure~\ref{IJK}, the AMES-Dusty models reproduce
well the locus  of the coolest dwarfs which deviate  from that of main
sequence  stars red  values of  $J-K$ as  dust effects  grow  in their
atmospheres  with decreasing effective  temperature (see  also Leggett
\etal\  1998\nocite{leg98}  for  more  such  color  comparisons).   It
appears, therefore, that this full-dusty limit where grain settling is
negligible is adequate to reproduce the global properties of late-type
low-mass  stars  and young  or  massive  brown  dwarfs with  $\teff\le
1800$K.  Below  this temperature, the AMES-Dusty  models keeps getting
redder  and do  not correspond  to the  properties of  known  T dwarfs
illustrated in this diagram by the position of Gl229B and SDSS1624.

The  locus of the  AMES-Cond models  for their  part depends  upon two
major uncertainties.  The first, likely  tied to the second, is a hump
of flux excess  between 0.8 and 0.93~$\mu$m, i.e.   in the I-bandpass,
which prevent  the Cond models  to become redder than  $I-J=4.2$.  The
second is the description of the  far wings of the absorption lines of
K~I and Na~I~D as discussed above and illustrated in Figure \ref{vdw}.
In Figure~\ref{IJK}  we show  two grids of  Cond models:  one computed
with a coverage of the line wings opacity contributions of 5000\AA\ on
each  side of  each atomic  line core  (long dashed  line),  the other
computed  with a maximum  coverage of  15000\AA\ (short  dashed line).
Both grids use Lorentz profiles  for the atomic lines.  Obviously, the
profile of the optical Na~I~D  and K~I doublets is no longer Lorenzian
beyond  5000\AA\ from  the  line cores  as  also been  noted found  by
\cite{BMS00}.  Since T dwarfs appear in the cone defined by these Cond
models, an adequate  theory of line broadening could  be sufficient to
reproduce their properties.  Yet no theory exists for the treatment of
the far  wings of alkali  elements broadened by collisions  with H$_2$
and helium  species to this  date \cite[]{BMS00}.  Until  these become
available, the present grid with a line wing coverage of 5000\AA\ seem
to provide an acceptable compromise  and limiting case (with the Dusty
models) for the spectroscopic properties of brown dwarfs.

\section{\label{concl}Discussion and Conclusions}

We have investigated the two limiting  cases of dust in low mass stars
and  brown dwarfs  atmospheres by  comparing  two sets  of models:  1)
AMES-Cond  with models ranging  from $\teff  = 3000$  to 100K  in 100K
step, and with gravities ranging from $\log g= 2.5$ to 6.0 in steps of
0.5  dex at  solar metallicity,  and  2) AMES-Dusty  with models  from
$\teff =  3000$ to 1400K in  steps of 100K, with  gravity ranging from
$\log  g=  3.5$ to  6.0  in  steps of  0.5  dex.   The AMES-Cond  grid
corresponds  to the  case  where  all dust  has  disappeared from  the
atmosphere  by  gravitational   settling  while  the  AMES-Dusty  grid
describes   the   case   of   negligible   settling   throughout   the
atmosphere. The  two sets of models  rely on the  assumption that dust
forms in equilibrium  with the gas phase.  See  also \cite{CBAH00} for
the corresponding evolution models of late type stars and brown dwarfs
and a more detailed description of their photometric properties.

From  the comparison  of  the  two limits  described  here, and  their
comparisons to  observations (see  Section \ref{color}), we  find that
dwarfs  with  $\teff \ge  1800$K  are  successfully  described by  the
full-dusty limit and  conclude that dust must be  in close equilibrium
with  the   gas  phase  with  little   sedimentation,  or  compensated
sedimentation  effects.   This would  be  the  case  if, for  example,
convective  mixing was  efficient in  returning material  to  the line
forming  regions.    We  are   exploring  these  issue   by  modeling,
hydrodynamically, the convection in  3D models (Ludwig \etal\ 2000, in
preparation).

Observations  of T  dwarfs such  as  Gliese 229B  indicate that  brown
dwarfs with  $\teff \le  1300$K are, on  the other hand,  more closely
following the  full-settling limiting case  as has been shown  also by
\cite{tsuji96a,tsuji96b,gl229b} and \cite{marley96}. However there has
been a history of failure to explain the optical spectra of these cool
brown   dwarfs  since   the   discovery  of   Gl229B.   For   example,
\cite{Griffith98} claimed  a veiling due to  photo-dissociation by the
parent  starlight  in the  upper  brown  dwarf  atmosphere.  But  this
suggestion did not pass the acid test, and free-floating T dwarfs were
discovered               \cite[]{Tsvetanov00}.                Finally,
\cite{tinney98,gd165b,BMS00}  and \cite{liebert00}  have  stressed the
importance     of     the     $\lambda$5891,5897\AA\    Na~I~D     and
$\lambda$7687,7701\AA\ K~I  resonance doublets.  Our  analysis recover
the latter results in predicting the  red wings of these lines to heat
up  the photospheric  thermal structure  by as  much as  100K,  and to
define the  pseudo-continuum out to  at least 1.1~$\mu$m in  T dwarfs.
This indicates  that red flux out  to 1.1~$\mu$m must  be sensitive to
gravity.  We also  find that the traditional Lorentz  profile does not
describe  adequately their red  wings, beyond  5000\AA\ from  the line
center:  it overestimate  the wing  opacity contribution.   While this
overestimation has negligible effects on the thermal structure as long
as  the lines  are included  in the  calculation, it  is  important to
realize  that an adequate  handling of  the collisional  broadening of
alkali  lines by  H$_2$ and  He must  be developed  in order  to model
successfully  the optical  to 1.1~$\mu$m  flux and  $VRI$ colors  of T
dwarfs  \cite[]{BMS00}.  Nevertheless  we believe  that  our AMES-Cond
models,   limited  in  the   coveraged  of   the  atomic   line  wings
contributions,  can provide  the  full-settling limit  that they  were
intended to define.

Based on the  Cond models, we also find that  currently known T dwarfs
with effective temperatures around 1000K are in a special regime where
several of  their important spectral  features start to  reverse their
behavior with  decreasing $\teff$: Cs~I  lines weaken as Cs  begins to
lock into CsCl,  and CO and NH3  bands begin to grow in  strength as a
result  of  decreasing  background  opacities.   While  \cite{Noll97},
\cite{Griffith00},  and \cite{saumon00}  all fail  to  reproduce these
features  in Gl229B  with  solar composition  models in  thermodynamic
equilibrium, it is perhaps due to the fact that each of these spectral
synthesis analysis were based upon thermal structures exempt of atomic
line  opacities?   Indeed, in  the  fit  to  the overall  spectrum,  a
sub-solar composition could compensate (via an increased pressure) for
the  neglect  of  atomic  line  opacities  in  the  thermal  structure
calculations.  To clarify  the issue of the metallicity  of Gl229B, we
need to explore the composition  of the parent star Gl229A.  This work
is in progress  and will be published shortly.   However, we feel that
it is unlikely,  though not impossible, that the  first known T dwarfs
be metal-deficient, while more massive  brown dwarfs are only found in
metal-rich environments.

There  remains a  regime between  1300  and 1800K  where brown  dwarfs
should  behave between these  limits.  We  hope that  our two  sets of
models can help bracket brown dwarf's properties and identify objects.
They can also be used to evaluate the effects of intrinsic variability
as  weathering effects  that  cause  clouds to  form  and vanish.  The
full-dusty models  give the  aspect of a  cloudy atmosphere  while the
Cond models resemble a cloud-free sky.

In the  atmospheres of brown dwarfs  and late-type M  dwarfs, the dust
likely  forms in  clouds distributed  more or  less evenly  across the
dwarf's surface as  we observe on planets.  These  cloud layers should
be  well-confined  close  to   the  deepest/hottest  level  where  the
grain-type         can          condense         \cite[see         for
example]{lunine89,tsuji99,Ack00}.   But   the  physical  process  that
defines this confinement of the  cloud layers are not known.  It could
either  be: i)an  inefficient condensation  of the  dust in  the upper
atmosphere, or ii) an efficient gravitational settling (sedimentation)
of  the  dust in  those  upper  layers.   Time-dependent grain  growth
analysis  must be done  to determine  the first.   This work  is under
progress at  the Berlin University.  But our  CE calculations indicate
that the local [T,P] conditions  favor dust formation.  The second can
be  understood by  opposing sedimentation  and convective  mixing, one
pushing  the  grains down,  the  other  bringing  upwards material  to
condense.  Since  the convection zone retreats  progressively from the
photosphere  with  decreasing  $\teff$,  it  seems  to  be  a  natural
explanation of the  fact that dust also retreats  from the photosphere
as $\teff$ decreases.  This is why we mention the potential importance
of gravitational settling throughout this paper.

The Cond models represent a limiting case where these cloud layers are
all sitting below the photosphere,  independently of the cause for the
cloud confinement. The Dusty models on the other hand represent a case
where  confinement does  not occur.   So it  is clear  that  nature is
likely between  the Cond and Dusty limits,  with partial gravitational
settling  occurring with  a fraction  that  varies with  depth in  the
atmosphere,  the   detailed  modeling   of  this  process   relies  on
characteristic  diffusion time  scales for  several processes  such as
condensation, sedimentation, coagulation and convective mixing to name
a few which are not known  accurately for the type of grains important
under the  conditions prevailing  in brown dwarf  atmospheres.  Models
incorporating these  effects can  therefore only be  exploratory.  The
two  limits described  here will  remain useful  until the  physics of
these processes become solidly mastered.

All the models discussed in this paper are available upon request.  We
will also provide, as we have  in the past, colors computed from these
spectra on any requested color  system. Please send requests to France
Allard and consult the CRAL anonymous ftp site.

\acknowledgments

We wish to thank specially Richard Freedman (NASA-Ames) for joining us
in the analysis of the VO and CrH line formation in M dwarfs and brown
dwarfs,  Hans  G.   Ludwig  (Lund)  for  a  lot  of  very  instructive
discussions and  for his interest  in dust formation in  brown dwarfs,
and Tristan  Guillot (Obsv. de  Nice) for his support  and interesting
collaborations  to  come.  We  thank  also  Gilles Chabrier,  Isabelle
Baraffe  and  Travis  Barman   for  proofreading  and  providing  some
orientation to the draft.  This  research is supported by CNRS as well
as  NASA LTSA  NAG5-3435  and a  NASA  EPSCoR grant  to Wichita  State
University.   Peter  Hauschildt  and  Andreas  Schweitzer  acknowledge
support in  part from NASA ATP  grant NAG 5-3018, NAG  5-8425 and LTSA
grant  NAG  5-3619  to  the   University  of  Georgia.   Some  of  the
calculations presented in this paper  were performed on the IBM SP2 of
the  CINES, and the  UGA UCNS  at the  San Diego  Supercomputer Center
(SDSC)  and the  Cornell Theory  Center (CTC),  with support  from the
National Science  Foundation.  We thank  all these institutions  for a
generous allocation of computer time.

\clearpage

\bibliography{}

\clearpage

\centerline{\bf Figure captions }

\clearpage

\end{document}